 \documentclass[]{aa} 
\usepackage{graphicx}
\usepackage{lscape}
\usepackage{rotating}
\usepackage{natbib}
\usepackage{txfonts}
\newcommand{\ome}{$\omega$~Cen}

\def\logg{\mbox{log~{\it g}}}

\def\teff{\mbox{$T_{\rm eff}$}}

\def\rpro{\mbox{$r$-process}}
\def\spro{\mbox{$s$-process}}
\def\ncap{\mbox{$n$-capture}}

\begin{document}
   \title{Chemical enrichment mechanisms in Omega Centauri: \\
          clues from neutron-capture elements\thanks{Based 
          on data obtained with the ESO UVES spectrograph 
          during the observing programmes 165.L-0263 and 67.D-0245}
 }


   \author{V. D'Orazi
          \inst{1}
          \and
	  R.G. Gratton
          \inst{1}
	  \and
          E. Pancino \inst{2}
          \and
	  A. Bragaglia \inst{2}
          \and
          E. Carretta \inst{2}
	  \and
	  S. Lucatello \inst{1}
          \and
	  C. Sneden \inst{3}
	   }

   \institute{INAF $-$ Osservatorio Astronomico di Padova, 
              vicolo dell'Osservatorio 5 , I-35122, Padova, Italy \\
              \email{valentina.dorazi@oapd.inaf.it, 
              raffaele.gratton@oapd.inaf.it, sara.lucatello@oapd.inaf.it}\\
              \and 
              INAF $-$ Osservatorio Astronomico di Bologna, 
	      via Ranzani 1, I-40127, Bologna, Italy \\
	      \email{elena.pancino@oabo.inaf.it, 
              angela.bragaglia@oabo.inaf.it, eugenio.carretta@oabo.inaf.it}\\
	      \and 
	      Department of Astronomy and McDonald Observatory, 
	      The University of Texas, Austin, TX 78712, USA \\ 
	      \email{chris@verdi.as.utexas.edu} 
	       }

   \date{Received:   Accepted: }

 
  \abstract
   {In the complex picture of multiple stellar populations in globular 
    clusters (GCs), a special role is played by NGC~5139 
    ($\omega$~Centauri). 
    At variance with the majority of GCs, \ome~ exhibits significant 
    star-to-star variations in metallicity and in 
    relative neutron-capture element abundance ratios with respect
    to Fe, along with split evolutionary sequences as revealed from 
    colour-magnitude diagrams. 
    Combining information from photometry and spectroscopy, 
    several studies suggested that an age spread of several Gyr has to 
    be invoked to explain (at least partially) some of the observed features. 
    However, a comprehensive understanding of the formation, evolution 
    and chemical enrichment processes is still not at hand. }
   {Relatively metal-rich \ome\ stars display neutron-capture abundance
    distributions dominated by contributions from the \spro, but it
    is not clear what roles have been played by the so-called $main$ 
    and $weak$ \spro\ components in generating these abundances. 
    To gain better insight into this question we derived lead 
    (Pb) abundances for several \ome\ cluster members, because this 
    element can only be produced by the $main$ \spro.}
   {We analysed high-resolution UVES@VLT spectra of a sample of twelve 
    red-giant branch stars, deriving abundances of Pb and also of Y, 
    Zr, La, Ce, Eu, and the C+N+O sum. 
    Spectral synthesis was applied to all features, taking into 
    account isotopic shifts and/or hyperfine structure as needed.}
   {We measured for the first time the Pb content in \ome, discovering 
    a clear hint for a Pb production occurring at [Fe/H]$>$$-$1.7 dex. 
    Our data suggest that the role of the $weak$ component in the 
    production of s-process elements is negligible. 
    Moreover, evidence gathered from the abundances of other elements 
    indicates that the $main$ component occurring in this GC is peculiar 
    and shifted towards higher mass polluters than the standard one.
   }  
  {}

  \titlerunning{Heavy-elements in $\omega$ Centauri}
   \authorrunning{V. D'Orazi et al.}             
\maketitle
\section{Introduction}
About 30 years ago new efficient medium-high resolution spectrometers
began appearing on very large telescopes, allowing the acquisition 
of high-quality, high S/N spectra for large samples of globular
cluster (GC) stars. 
This new technology led to extensive spectroscopic surveys that
substantially modified the classical view of cluster birth and evolution.
For a couple of decades now we have understood that many (all?) GCs 
exhibit significant star-to-star variations in their light-element 
contents (see Kraft 1994\nocite{kr94}; 
Gratton et al. 2004\nocite{gr04}; Martell 2011 for reviews of these discoveries).

Photometric colour-magnitude observations have revealed 
split evolutionary sequences in a handful of GCs (e.g. Lee et al. 1999; 
Pancino et al. 2000; Bedin et al. 2004; Milone et al. 2008; Han et al. 2009).
But it has taken high-resolution spectroscopy to truly establish 
the complex nature of GCs.
With their peculiar abundance pattern, GCs are fundamentally different 
from the other Galactic stellar populations (young/old open clusters, 
see e.g., Gratton et al. 2004\nocite{gr04}; 
de Silva et al. 2009\nocite{ds09}; Pancino et al. 2010a\nocite{p10a}; Bragaglia et al. 2011, submitted).

Analysing a sample of more than 1500 giants in 19 GCs,
Carretta et al. (2009a,b\nocite{c09a,c09a}) concluded that all the surveyed GCs, 
although characterised by a homogeneous composition in iron-peak 
and the heavy $\alpha$-elements (e.g., Ca and Ti), present 
variations in light elements.
These variations do not occur randomly:  the abundances of 
CH$-$CN, O$-$Na, and Mg$-$Al are 
anticorrelated with each other in stars at all evolutionary stages 
(e.g., Cohen 1999; Kayser et al. 2008\nocite{kay08}; Pancino et al. 2010b\nocite{p10b}; 
Carretta et al. 2009a,b\nocite{c09a,c09b}; Lind et al. 2009\nocite{li09}).
Globular clusters contain stars with high O (Mg, and C) and low Na 
(Al, and N) abundances, which constitute the primordial 
cluster population (first-generation stars), and stars with
considerable enhancement in Na, Al, and N accompanied by
depletion in O (Mg, and C), which are second-generation stars.
The light-element abundance pattern of the second-generation
stars (which can be seen in stars of all evolutionary stages)
can be identified with hot H-burning nucleosynthesis (Denisenkov \& Denisenkova 1989; Langer et al. 1993).
However, the nature of element-donating stars of the previous 
generation, whose ashes provided the material from which 
second-generation stars formed, is still uncertain.
Various scenarios, none universally-accepted yet, have been 
proposed:  intermediate-mass AGB stars (Ventura et al. 2001); 
fast rotating massive stars (Decressin et al. 2007\nocite{dec}); 
and massive binaries (de Mink et al. 2009\nocite{dmink}).

The abundances of neutron-capture elements (\ncap, Z $>$~30) 
generally show little star-to-star scatter in most GCs (Armosky et al. 1994; James et al. 2004).
Typically an \ncap\ distribution more heavily weighted towards
the \rpro\footnote{
Elements heavier than iron are mainly synthesised in neutron-capture 
reactions, which can be divided into s(low)-process and r(apid)-process, 
where slow and rapid refers to the $\beta$-decay timescale.}
than the \spro\ is apparent in GCs, most easily seen in ratios 
[Eu/Ba]~$\approx$~$+$0.4 (as summarized by Gratton et al. 
2004\nocite{gr04}).
Recently, D'Orazi et al. (2010)\nocite{d10} have shown that the 
\spro-dominated element barium (Ba) is mostly characterized by 
a single abundance value within individual GCs
(see however Roederer 2011 for the spread in $r$-process elements within some GCs).

In addition to the ``prototypical" GCs, like NGC~6752 (see Carretta 
et al. 2009b\nocite{c09b}; Pasquini et al. 2005\nocite{p05}; 
Shen et al. 2010\nocite{s10}), the large family of GCs includes 
some peculiar cases. 
The very metal-poor GC M15 (NGC~7078) has large star-to-star \rpro\ scatter
(e.g. $\delta$[Eu/Fe]~$\approx$~0.5; Sneden et al. 1997\nocite{sn97}, 
Sobeck et al.  2011\nocite{so11}).
In contrast, M22 (NGC~6656) exhibits an overall metallicity variation 
($\delta$[Fe/H]~$\sim$~0.15) that is positively correlated with 
the variations in $s$-process elements (e.g.  
$\delta$[Ba,Y,Zr/Fe]~$\approx$ 0.4; Marino et al. 2009\nocite{mar09}).
In NGC~1851, Yong and Grundahl (2008) found a scatter in [Fe/H] of 0.09 dex, 
while a slightly smaller variation (i.e., $\delta$[Fe/H]$\approx$0.06) has been detected by Carretta et al. (2010a). 
Ba seems instead to vary from a factor of four to more than 
one dex (Yong \& Grundahl 2008\nocite{yo08}, 
Villanova et al. 2010\nocite{vil10}, and Carretta et al. 2011).
Looking outside our Galaxy, M54 (NGC~6715), currently located 
in the centre of the Sagittarius dwarf spheroidal galaxy 
(e.g., Bellazzini et al. 2008\nocite{b08}), has been recently 
scrutinised by Carretta et al. (2010b). 
This work confirmed the presence of an internal iron spread, as 
previously suggested by e.g., Sarajedini \& Layden (1995\nocite{sl95}), 
of about 0.19 dex (r.m.s), with the bulk of stars at [Fe/H]$\sim-$1.6 
and a long tail extended at higher metallicity. Finally, in the bulge metal-poor GC NGC~6522 Barbuy et al. (2009) found a spread in the Ba content
of about $\sim$ 0.5 dex (rms) by analysing eight stars (see Chiappini et al. 2011 for a more detailed discussion).

However, the most intriguing and spectacular case of a complex GC is 
$\omega$ Centauri (NGC~5139). 
It is the most massive GC of our Galaxy, with a total mass of 
2.5$\times$10$^6$~M$_{\odot}$ (van de Ven et al. 2006\nocite{vdv}). 
It has been the object of many photometric investigations 
since the 1960s (e.g. Wooley 1966\nocite{w66}; Anderson 1998\nocite{an98};  
Pancino et al. 2000\nocite{p00}; Bedin et al. 2004\nocite{b04}; 
Ferraro et al. 2004\nocite{fer04}; 
Sollima et al. 2005a, 2007\nocite{sol05,sol07}; 
Bellini et al. 2010\nocite{be10}). 
All these works found the presence of split evolutionary sequences, 
from the main-sequence (MS), to the subgiant branch (SGB) all the way to the  red-giant branch (RGB) tip. 
At the same time, spectroscopic surveys revealed variations 
in [Fe/H] up to a factor of $\sim$~10 and a steep trend 
with Fe of \ncap\ elements La and Ba (e.g., 
Norris \& Da Costa 1995\nocite{nd95}; Smith et al. 2000\nocite{sm00}, 
hereafter S00).  

Recently Johnson \& Pilachowski (2010\nocite{jp10}, JP10) published 
an extensive spectroscopic study for a sample of more 
than 850 RGB stars in \ome, building up the most comprehensive
database of this kind available so far.
The authors confirmed that the metallicity ranges from 
[Fe/H]$\approx$$-$2 to [Fe/H]$\approx$$-$0.5, and that abundances for 
heavy $\alpha-$elements (e.g. Si, Ca, Ti) and Fe-peak ones (e.g. Sc, Ni) 
satisfactorily match a Type-II SN abundance pattern.
On the other hand, the light elements (O, Na, Al) present 
variations larger than 0.5 dex and exhibit the classical GC 
proton-capture anticorrelation trends at almost all metallicities 
(with the exception of the most metal-rich stars).
JP10 found a roughly constant [$\alpha$/Fe] ratio, which they 
argue rules out a significant contribution from Type-Ia 
SN (see however Pancino et al. 2002\nocite{p02}, 
Origlia et al. 2003\nocite{or03} for a different view).
Finally among the \ncap\ elements, [La/Fe] strongly
increases with [Fe/H], not matched by a similar 
trend in [Eu/Fe] (JP10), suggesting that at [Fe/H]$\gtrsim$$-$1.6, 
the $s$-process becomes the dominant \ncap\ production mechanism. 
This clearly implies that long-duration timescales ($\gtrsim$1 Gyr) 
are required for the evolution of low-mass AGB, where the 
{\it main} component of the $s$-process production occurs.

Similar conclusions were also drawn by Marino et al. (2011\nocite{mar11}, 
M11), who analysed more than 300 \ome~giants. 
In trying to solve the problems of timescales in the evolution
of various abundance signatures, M11 speculated that the so-called {\it weak} 
\spro\ component (see e.g. Raiteri et al. 1993\nocite{rai93}; 
Travaglio et al. 2004\nocite{tr04}) might perhaps be responsible for the
enhancement in \spro\ elements. 
The {\it weak} component, whose source are probably massive AGB stars, 
is responsible for substantial amounts of the solar-system
\spro\ isotopes up to A~=~90 but contributes vanishingly small
amounts of heavier material.
The so-called {\it main} component accounts for the heaviest 
\spro\ element abundances (i.e. beyond Zr), and its production
site has been ascribed to thermally pulsing low-mass AGBs 
(M/M$_{\odot}$ $\sim$ 1$-$3).
Marino et al. suggested that at the lower metallicities
characteristic of \ome,  the {\it weak} component 
might produce heavier $n$-capture elements, up to Ba and La, owing to 
the higher neutron over iron-seed ratio.

To test this prediction, and in general to observationally 
constrain element enrichment timescales, we have conducted
a new high-resolution spectroscopy study of 12 \ome\ RGB stars
in the rarely-studied blue spectral region.
For this stellar sample we have derived abundances for Y, Zr, 
La, Ce, and Eu, and estimated the C+N+O abundance sums.
More importantly, we have derived for the first time lead 
(Pb) abundances in \ome. 
In the \spro, Pb can only be produced by the {\it main} component.
Detection of significant amounts of Pb in \ome\ giants indicates
that the {\it main} component has played a significant role in the production of the \spro\ elements in this GC.

The paper is organised as follows: in Sect.~2 we give information 
on the sample stars, data reduction and abundance analysis.
Our results are illustrated in Sect.~3, while their scientific 
implications are discussed in Sect.~4.  
A summary is given in Sect.~5.


\section{Observations and analysis}

\subsection{Sample stars and data reduction}\label{sect:analysis}

We considered a sample of 12 \ome~ RGB stars, covering the whole 
metallicity range, i.e.  from [Fe/H]~$\approx$ $-$2 to 
[Fe/H]~$\approx$ $-$0.5.
In Table~\ref{t:phot} we present their basic data.
For each star, column 1 has the ``Leiden'' stellar designation 
(van Leeuwen et al.  2000\nocite{vl00});
column 2 has the older ``ROA'' number (Woolley 1966\nocite{w66});
columns 3$-$5 have broad-band photometry;
and columns 6$-$9 have model atmosphere parameters 
T$_{\rm eff}$, $\log{g}$, [Fe/H], and $\xi$ that we have adopted from JP10.
While our sample is small and biased towards high metallicity with 
respect to the bulk of \ome~stars, these targets cover an interesting 
range for studying the enrichment of \ncap\ elements.

\begin{table*}
\caption{Identification, photometry, and adopted parameters for our programme stars (see JP10).}\label{t:phot}
\begin{center}
\begin{tabular}{lcccccccc}
\hline\hline
Leiden & ROA  &  V      &  B$-$V  & K   &  T${\rm eff}$ & log$g$ & [Fe/H] & $\xi$ \\
       &      &         &         &	&    (K)	&	&	 & (km~s$^{-1}$) \\
       &      &         &         &	&		&       &	 &		  \\
\hline       
16015 &  ~213 & 12.127  &  1.122  &  ~9.210 &  4510     & 1.05     & -1.92    & 2.00  \\
37247 &  ~238 & 12.430  &  1.163  &  ~9.363 &  4385     & 1.05	  & -1.88    & 1.75   \\
33011 &  ~159 & 11.879  &  1.337  &  ~8.715 &  4305     & 0.80	  & -1.75    & 2.05   \\
41039 &  ~256 & 12.251  &  1.230  &  ~9.190 &  4375     & 1.00	  & -1.73    & 1.95   \\
46092 &  ~~92 & 11.830  &  1.571  &  ~8.128 &  3990     & 0.45	  & -1.45    & 2.10   \\
34029 &  ~243 & 12.107  &  1.452  &  ~8.719 &  4170     & 0.80	  & -1.28    & 1.90   \\
44462 &  ~321 & 12.559  &  1.403  &  ~8.942 &  4040     & 0.90	  & -1.18    & 1.65   \\
60066 &  2118 & 13.086  &  1.253  &  10.330 &  4640     & 1.75	  & -0.98    & 1.55   \\
60073 &  ~211 & 12.266  &  1.613  &  ~8.525 &  3985     & 0.75	  & -0.82    & 2.00   \\
34180 &  ~517 & 13.030  &  1.503  &  ~9.329 &  4015     & 1.05	  & -0.79    & 1.95   \\
48323 &  ~500 & 13.081  &  1.461  &  ~9.273 &  3945     & 0.65	  & -0.73    & 2.00   \\
54022 &  2594 & 13.360  &  1.412  &  ~9.910 &  4135     & 1.15	  & -0.46    & 1.80   \\
\hline\hline
\end{tabular}
\end{center}
\end{table*}

One of the stars presented in this paper, ROA~211 
(Woolley 1966\nocite{w66}), or Leiden \#60073 
(van Leeuwen et al. 2000\nocite{vl00}), has already been studied by
Pancino et al. (2002\nocite{p02}). 
Its spectrum was obtained in June 2000 with the UVES spectrograph 
mounted at the ESO VLT, Paranal observatory, Chile. 
We used the 1$\arcsec$-wide slit; the spectral resolving power was
R~$\equiv$~$\lambda/\delta\lambda$~$\approx$ 45\,000. Signal-to-noise ratios 
-- per pixel -- near 6000~\AA\  are higher than 100, while around the Pb~{\sc i} lines 
at 3683 \AA~ and 4058 \AA~ we have S/N$\sim$20 and 50, respectively. 
All the other stars were obtained in a subsequent observing
run in April 2001, using a similar instrumental setup, which provided
a similar resolution and S/N ratio. 
A preliminary abundance analysis of these data was published 
by Pancino et al. (2003). 
The wavelength window for all our sample stars covers
3600~\AA~$\leq$ $\lambda$~$\leq$ 4600~\AA; only for the stars 
\#60073 and \#60066 were spectra available also at redder wavelengths 
(5400~\AA~$\leq$ $\lambda$~$\leq$ 8900~\AA).

The instrumental fingerprint was removed from the observed frames with
IRAF\footnote{IRAF is the Image Reduction and Analysis Facility, a
general purpose software system for the reduction and analysis of
astronomical data. IRAF is written and supported by the IRAF programming
group at the National Optical Astronomy Observatories (NOAO) in Tucson,
Arizona. NOAO is operated by the Association of Universities for
Research in Astronomy (AURA), Inc. under cooperative agreement with the
National Science Foundation.}, 
and the one-dimensional spectra were extracted, wavelength-calibrated, and roughly normalized to the continuum with IRAF tasks 
within the {\em echelle} package.

\subsection{Abundance analysis technique and line lists}\label{sec:abu}

The blue spectra of cool, metal-rich RGB stars are very 
crowded, and most lines are blended.
Therefore we employed a spectral synthesis analysis to derive 
all abundances in this study, using the ROSA code 
(Gratton 1988\nocite{gr88}) and the Kurucz (1993\nocite{kur93}) 
model atmospheres with the overshooting option switched on.
We adopted the model atmosphere parameters of JP10
for the programme stars (Table~\ref{t:phot}).
These model parameters were derived by JP10 in a uniform manner;
and uncertainties in these parameters are not important to the 
conclusions of our study (see below).

We examined several lists of potentially useful transitions for 
\ncap\ elements. 
Most of the possible lines proved to be too blended in our 
\ome\ spectra to give reliable abundance information.
In the end, our analysis rests on only a limited number of spectral 
features which we deem best for our purposes.
Here we provide brief comments on some of the transitions. 

\begin{table}
\begin{center}
\caption{Line atomic parameters and references} \label{t:atoms}
\begin{tabular}{lccr}
\hline\hline
 Ion                 &  $\lambda$  &  log$gf$  & Reference \\
                     &            &           &                     \\
\hline
Y~{\sc ii}           &  4398.02     & $-$1.00  &  Hannaford et al. (1982) \\
Zr~{\sc ii}          &  4208.98     & $-$0.51  &  Ljung et al. (2006) \\
La~{\sc ii}          &  4086.71     & $-$0.07  &  Lawler et al. (2001a)  \\
                     &  4322.50     & $-$0.93  &  Lawler et al. (2001)   \\
Ce~{\sc ii}          &  4073.47     &    0.21  &   Lawler et al. (2009)  \\
                     &  4349.79     &  $-$0.32 &   Lawler et al. (2009)  \\
Eu~{\sc ii}          &  4129.72     &    0.22  &   Lawler et al. (2001b)     \\
Pb~{\sc i}           &  3683.46     & $-$0.42  &  Rose \& Granath (1932)    \\
                     &  4057.82     & $-$0.20  &  Aoki et al. (2002)      \\
\hline
\end{tabular}
\end{center}
\end{table}

To determine the Pb~{\sc i} abundances we used the lines at 3683 and 4058~\AA. 
For the 3683~\AA\ line we adopted the hyperfine structure (hfs)
components from Rose \& Granath (1932\nocite{rg32}), 
and for the 4058~\AA\ line we adopted the hfs data given by 
Aoki et al. (2002\nocite{aok02}).
As is well known, the region around the 4058~\AA\ line is 
crowded with CH molecular lines. 
Therefore, prior to synthesizing this feature we estimated the 
strength of the CH contaminating features
using CH lines in the G-band spectral region (see below). 
We made an ultimately fruitless search for additional Pb~{\sc i} features,
using the line list from Bi{\'e}mont et al. (2000\nocite{bie00});
all possible transitions were either too weak, too blended, or both.

Y~{\sc ii} and Zr~{\sc ii} abundances were obtained from 
lines at 4398~\AA\ and 4208 \AA, respectively. 
The log$gf$ values were taken from Hannaford et al. 
(1982\nocite{han82}) for Y~{\sc ii} and 
from Ljung et al. (2006\nocite{lju06}) for Zr~{\sc ii}.
No isotopic shifts and/or hyperfine
data are available for those lines, but their effects should be almost 
negligible (e.g., Mashonkina et al. 2007\nocite{masho}).
\footnote{We could not derive Sr abundances because Sr~{\sc ii} 
features are too heavily saturated in the spectra of our cool red 
giant sample stars,  and our spectra do not include the 
Sr~~{\sc i} line at 4607 \AA.}

We analysed two features 
each for La~{\sc ii} and Ce~{\sc ii}: 4087 and 4322~\AA\ for La~{\sc ii} 
and 4073, 4350 \AA~ for Ce~{\sc ii}. 
For both La lines we adopted $gf$-values and hfs data 
presented by Lawler et al. (2001a\nocite{l01a}).
Additional La lines at 3988, 3995, and 4123 ~\AA\ can be seen on our
spectra, but they proved to be too blended for accurate abundances.
For the Ce lines we used the $gf$'s from Lawler et al. (2009\nocite{la09}).
Note that the dominant isotopes of Ce are even-Z, even-mass nuclei,
and so Ce~{\sc ii} lines are not affected by hyperfine splitting, and 
they also have negligible isotopic shifts.
Average abundances from individual La and Ce were computed, weighting 
the means by the individual line uncertainties.

The Eu~{\sc ii} abundances were derived from the line at 4129~\AA,
taking into account the strong hyperfine structures affecting
$^{151}$Eu and $^{153}$Eu odd isotopes; data for this transition
were taken from Lawler et al. (2001b\nocite{l01b}). 
As in the case of Pb, we made a special effort to identify more Eu~{\sc ii}
lines for analysis.
We did not find additional trustworthy transitions.
In particular: (a) the 4205~\AA\ line that is often used in Eu abundance
analyses was too contaminated by other species; (b) the 3907~\AA\
line was in a forest of other features that became far too strong at
higher \ome\ metallicities; 
(c) the 4435~\AA\ line was of reasonable
strength and always detectable, but far too blended with neighbouring
Ca~{\sc i} lines to yield good abundances,
and (d) the much weaker 
6645~\AA\ line, which would be ideal in our study, was unfortunately 
not part of our spectral coverage.
We summarise in Table~\ref{t:atoms} information on the employed atomic 
line lists with the corresponding oscillator strengths and references.\

As mentioned above, lines of the CH G-band at 4300 \AA~ were used 
to derive the C abundances. 
We adopted as initial $gf$~values the ones provided by 
B. Plez (private communication).
We then altered the band oscillator strengths to yield an acceptable solar C abundance,
log$\epsilon$(C)$_\odot$=8.54.
The CN bandhead at 4215 \AA~ was synthesised to obtain the N content, 
employing the line list given by Gratton (1985) 
and using the C values just derived; we assumed 
for the Sun an oxygen abundance of log$\epsilon$(O)$_{\odot}$=8.92.
\begin{center}
\begin{figure*}[htbp]
\includegraphics[width=17cm]{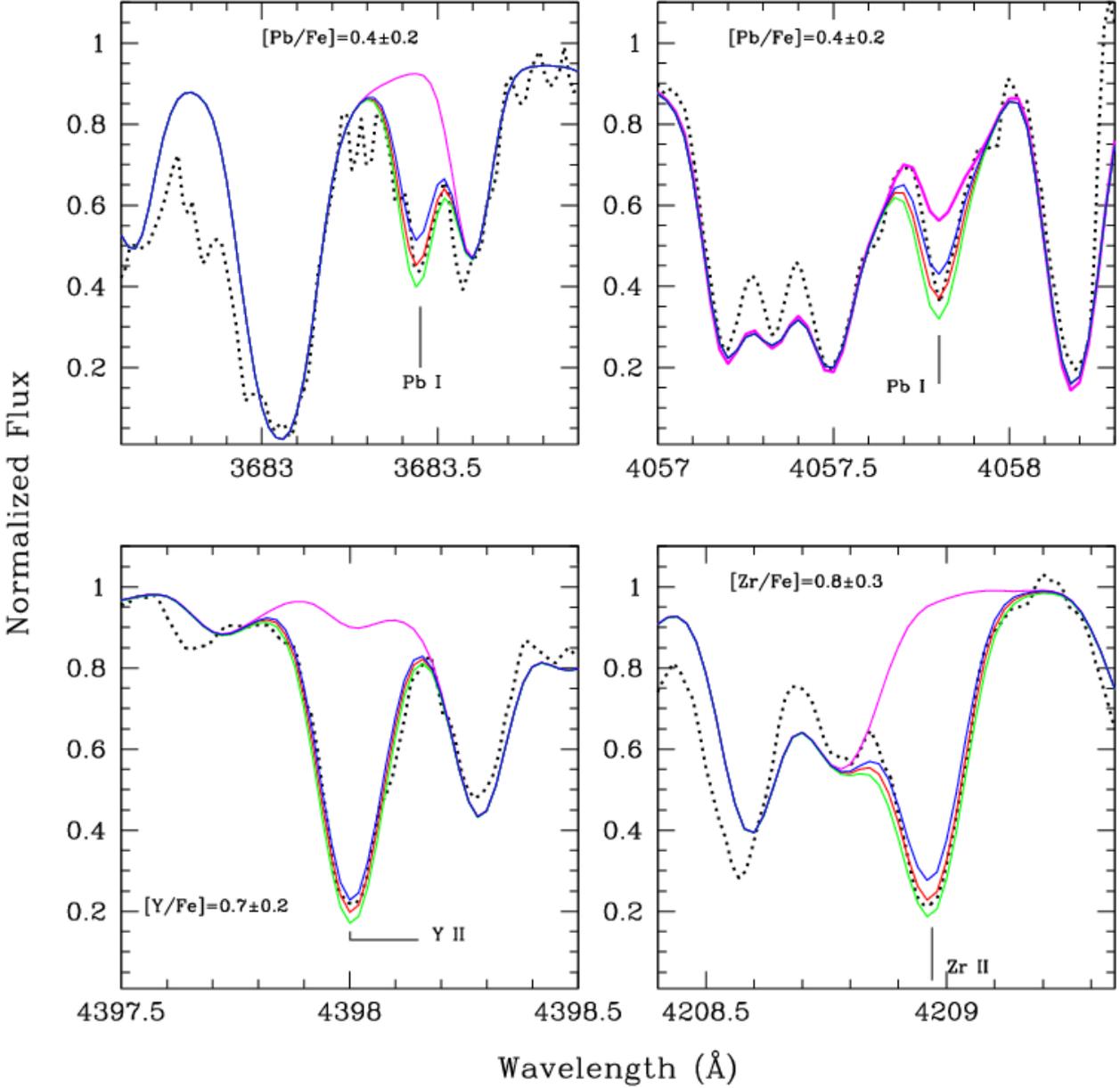}

\caption{Observed spectrum (dotted lines) for the star 
         \#34029 and spectral syntheses for Pb, Y, and Zr.
         Best fit (red lines) and errors (blue and green lines) are shown, 
	 along with syntheses with
	 log$\epsilon$(X)$\approx$$-$$\infty$ (magenta).} \label{f:obs}
\end{figure*}
\end{center} 
\begin{center}
\begin{figure*}[htbp]
\includegraphics[width=17cm]{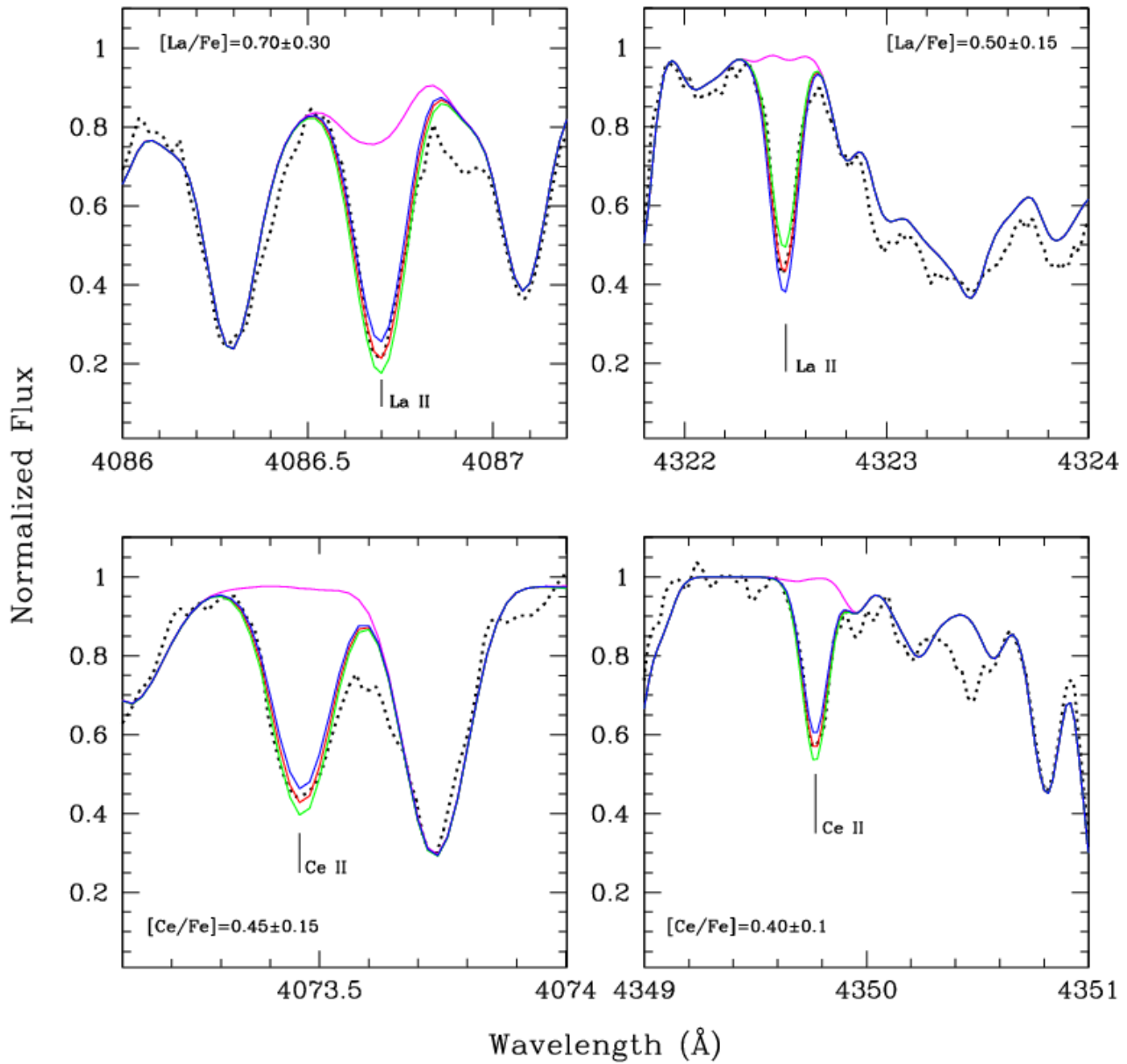}
\caption{Same as Figure~\ref{f:obs} but for the La and Ce lines.}\label{f:obs2}
\end{figure*}
\end{center} 
\begin{center}
\begin{figure*}[htbp]
\includegraphics[width=17cm]{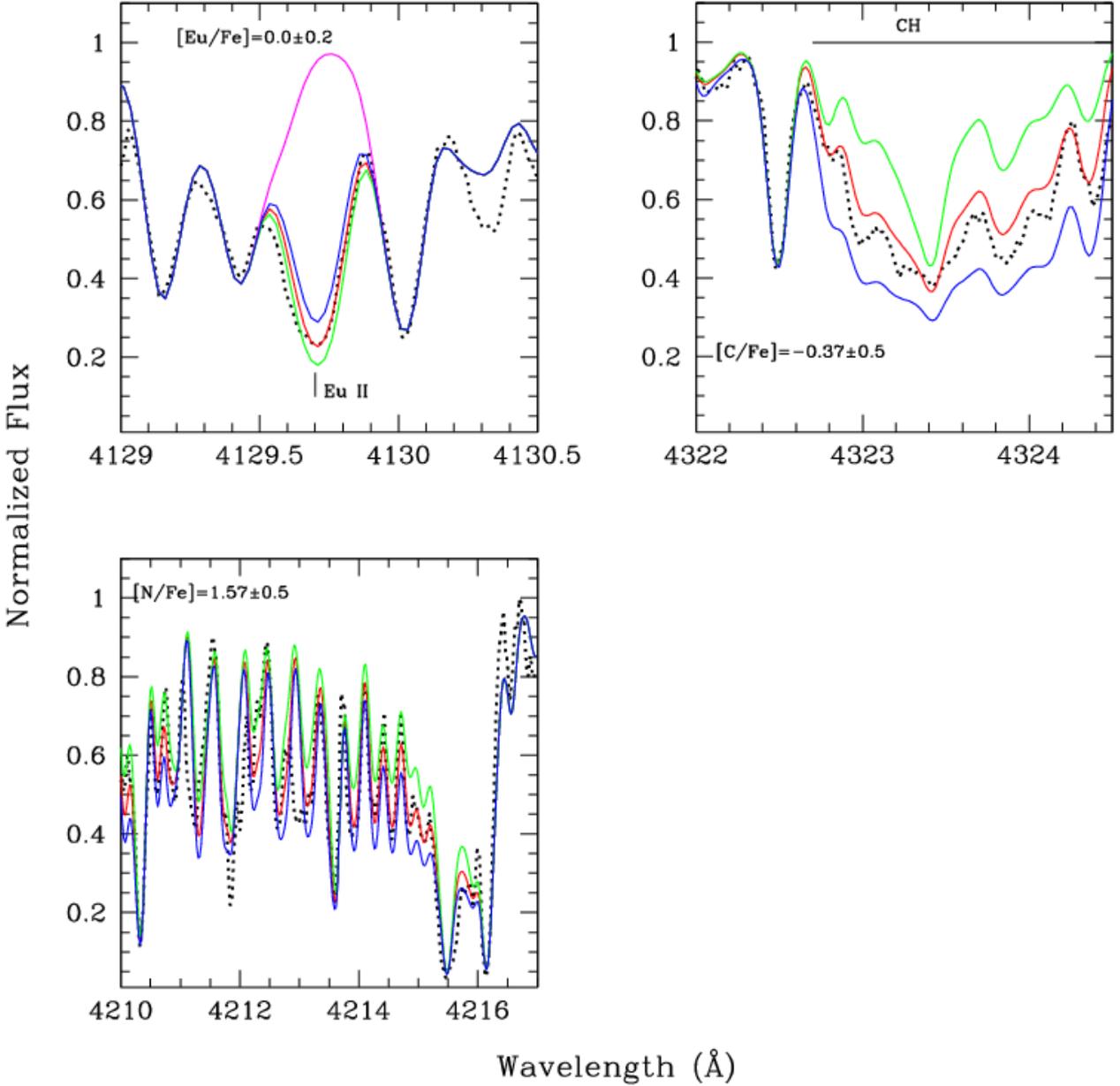}
\caption{Same as Figures~\ref{f:obs} and ~\ref{f:obs2}, but for the [Eu/Fe], CH and CN bands. 
         For the CN bandhead at 4215 a larger spectral window 
         is shown.}\label{f:obs3}
\end{figure*}
\end{center} 
\begin{table}
\caption{Abundances for the Sun and Arcturus.}\label{t:sun}
\begin{center}
\begin{tabular}{lccl}
\hline\hline
Element & log~$\epsilon$(Sun) & log~$\epsilon$(Arcturus) & log~$\epsilon$(Arcturus) \\
        &                     &                          &  ~~~~~~~~(S00) \\
\hline	                         
        &                     &       &      \\
C	     &	 8.54 &   8.05 &  ... \\
N	     &	 7.99 &   7.92 &  ...\\
Y~{\sc ii}   &	 2.24 &   1.74 & 1.40\\
Zr~{\sc ii}  &	 2.56 &   2.06 & 1.92 \\
La~{\sc ii}  &   1.22 &   0.52 & 0.62\\
Ce~{\sc ii}  &	 1.61 &   0.81 & 0.43 \\
Eu~{\sc ii}  &	 0.51 &   0.16 & 0.22 \\
Pb~{\sc i}   &	 2.05 &   1.12 &  ... \\
\hline\hline
\end{tabular}
\end{center}
\end{table}

\subsection{Optimisation and tests of the line lists}\label{sec:lintest}

As first step, we optimised our line lists for all transitions in the solar spectrum (Kurucz et al. 1985\nocite{kur85}); 
the adopted abundances for the Sun are listed in Table~\ref{t:sun}. 
Subsequently, we analysed the Arcturus ($\alpha$~Bootes) 
spectrum\footnote{Available at ftp://ftp.noao.edu/catalogs/arcturusatlas/}, 
given the similarities of this star's atmospheric parameters 
to those of our \ome~giants.
Adopted parameters for Arcturus were taken from Fulbright 
et al. (2007\nocite{ful07}), namely \teff~=~4280~K, \logg~=~1.55, 
$\xi$~=~1.61 kms$^{-1}$, and [Fe/H]~= $-$0.50.
Our derived abundances are reported in Column~3 of Table~\ref{t:sun}, 
where we also listed results from S00 (Column~4).

Referring to  Table~\ref{t:sun}, we derive for Arcturus [X/H]~= $-$0.5 for 
both Y and Zr, $-$0.60 for La, $-$0.80 for Ce, and $-$0.4 for Eu.
If we accept the Fulbright et al. (2007) [Fe/H] value, then 
[Y/Fe]~= [Zr/Fe]~= 0.0, [La/Fe]~= $-$0.1, [Ce/Fe]~= $-$0.3, and
[Eu/Fe]~= $+$0.1.
It is difficult to compare our abundances with those of major past 
analyses of Arcturus, because essentially all of them concentrated on
the yellow-red spectral regions, and because they were published
prior to the recent re-evaluation of transition data for many
of the \ncap\ species considered here.
With these cautions, note that some significant study-to-study
scatter has been reported.
For example, Norris \& Da Costa (1995) suggested that the Arcturus
relative abundance of all \ncap\ were essentially all the same, 
i.e. [\ncap/Fe]~= $-$0.3.
Peterson et al. (1993) derived [Y/H]~= $-$0.5 and [Zr/H]~= $-$0.6,
in good agreement with our values.
However, their abundances for the heavier \ncap\ elements average
about 0.3~dex larger than ours, as they obtained [La/H]~= $-$0.4, 
[Ce/H]~= $-$0.5, and [Eu/H]~= $-$0.1.
As a final example, S00 found lower values
than we did for the lighter \ncap\ abundances, [Y/H]~= $-$0.8 and 
[Zr/H]~= $-$0.6, but their heavier \ncap\ abundances were scattered
compared to ours, as they derived [La/H]~= $-$0.6, [Ce/H]~= $-$1.2, and 
[Eu/H]~= $-$0.3.
We conclude that our abundance set for Arcturus is reasonable compared
to previous work, and that the definitive study of \ncap\ abundances
in this star has yet to be conducted.

Assuming an [O/Fe]=+0.40 (e.g., Lambert \& Ries 1981\nocite{lr81};
Cottrell \& Sneden 1986\nocite{cs86}; Peterson et al. 1993), we confirmed 
that [C/Fe] ratio is approximately solar in Arcturus, as also found 
in these previous studies.
Moreover, our determination of [N/Fe]~= +0.4  is only slightly higher
than those values reported in the earlier work ([N/Fe]~= $+$0.2 to $+$0.3).

Abundances from spectral synthesis are mainly affected by two sources 
of uncertainties: errors owing  to the best-fit determination 
and to the stellar parameters, the last being retrieved 
from JP10 for all our sample stars. 
Because most of our features are very strong (near to the saturation 
part of the curve of growth), this is by far the predominant 
uncertainty, especially when we have only one feature for each element. 
These errors are reported in Table~\ref{t:abu}, where we show 
our final (average) abundances (see next section). 
When two spectral lines were available for a particular specie, we 
adopted the standard deviation from the mean of 
the two lines (rms) as an error estimate.  

Errors caused by stellar parameter
uncertainties are dominated by microturbulence ($\xi$) uncertainties,
because the spectral lines under scrutiny are usually very strong.
An error in T$_{\rm eff}$ of $\pm$100K and in $\log{g}$ of 
$\pm$0.30 dex (see JP10) results in a maximum variation of 
$\Delta$=0.15 dex for [X/Fe]. 
When we vary $\xi$ by 0.25 km~s$^{-1}$, this implies
a $\Delta$[X/Fe] of 0.25$-$0.30 dex (depending on the species).

In Figures~\ref{f:obs}$-$\ref{f:obs3}
examples of spectral synthesis are shown for star 
\#34029 for all transitions under analysis. 
As can be seen in the upper panels of Figure~\ref{f:obs}, while the 
Pb line at 4058~\AA~suffers from a non-negligible contribution 
from CH molecular features, the bluer line at 3683~\AA~ provides 
reliable information on the Pb content of this star. 
We obtain as best-fit value [Pb/Fe]=0.40$\pm$0.20. 
We also stress that taking into account the C abundance for this star
(see Table~\ref{t:abu}), we derived the 
same value of [Pb/Fe]=0.40$\pm$0.20 from the 4058~\AA~line.
In general we can conclude from our analysis that to derive accurate Pb measurements, 
the employment of the blue line at 3683~\AA~ is needed.
%
%
%
%
%
%

\section{Results}
\subsection{Neutron-capture elements}

Our derived abundances are summarised in Table~\ref{t:abu}, 
where we list in column~1 the Leiden stellar names, in columns 2$-$9 
the [X/Fe] ratios along with their internal uncertainties 
(see Section~\ref{sec:abu}), and in column 10 the CNO abundance sums.
As pointed out by JP10, and as evident in Tables~\ref{t:phot} and~\ref{t:abu}, 
there is a slight dependence of metallicity (and abundances 
in general) on {\bf T$_{\rm eff}$ values}, with the most metal-rich stars 
being mainly characterised by lower temperature values.
The average T$_{\rm eff}$ for stars with [Fe/H]$\leq$-1.28 is
4289 K (rms=184), while for the more metal-rich stars it is 
4127 K (rms=260), resulting in a difference of 162K.

\begin{figure*}[htbp]
\begin{center}
\includegraphics[width=15cm]{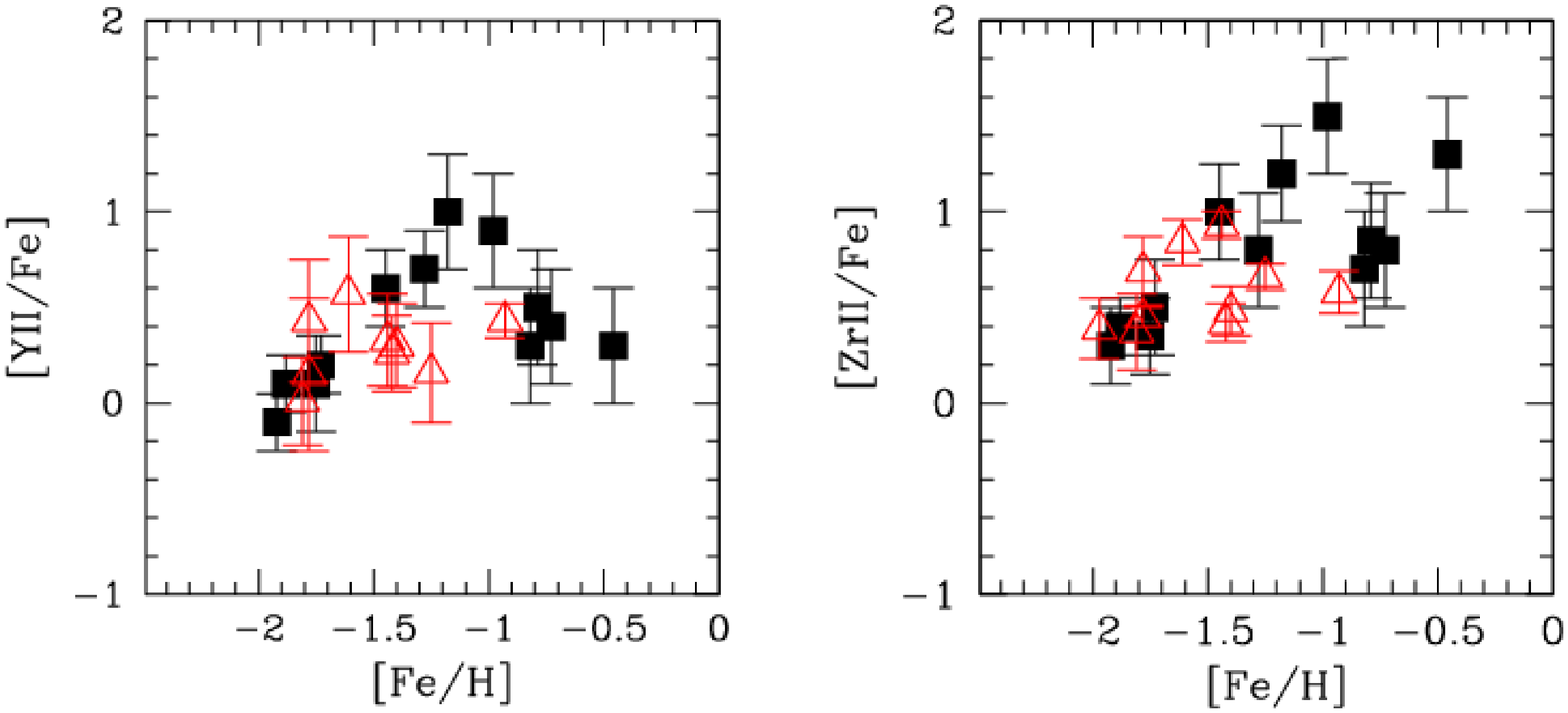}
\includegraphics[width=15cm]{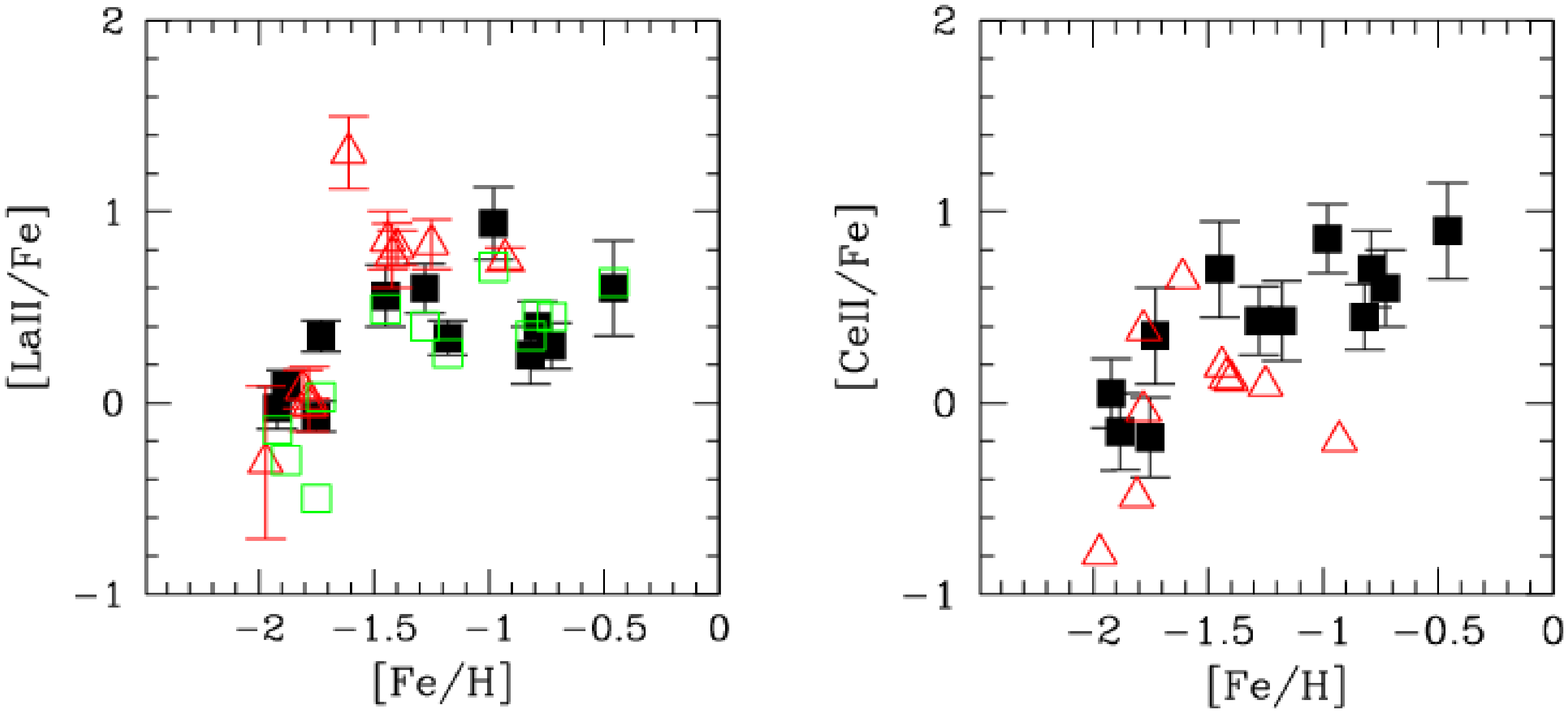}
\includegraphics[width=15cm]{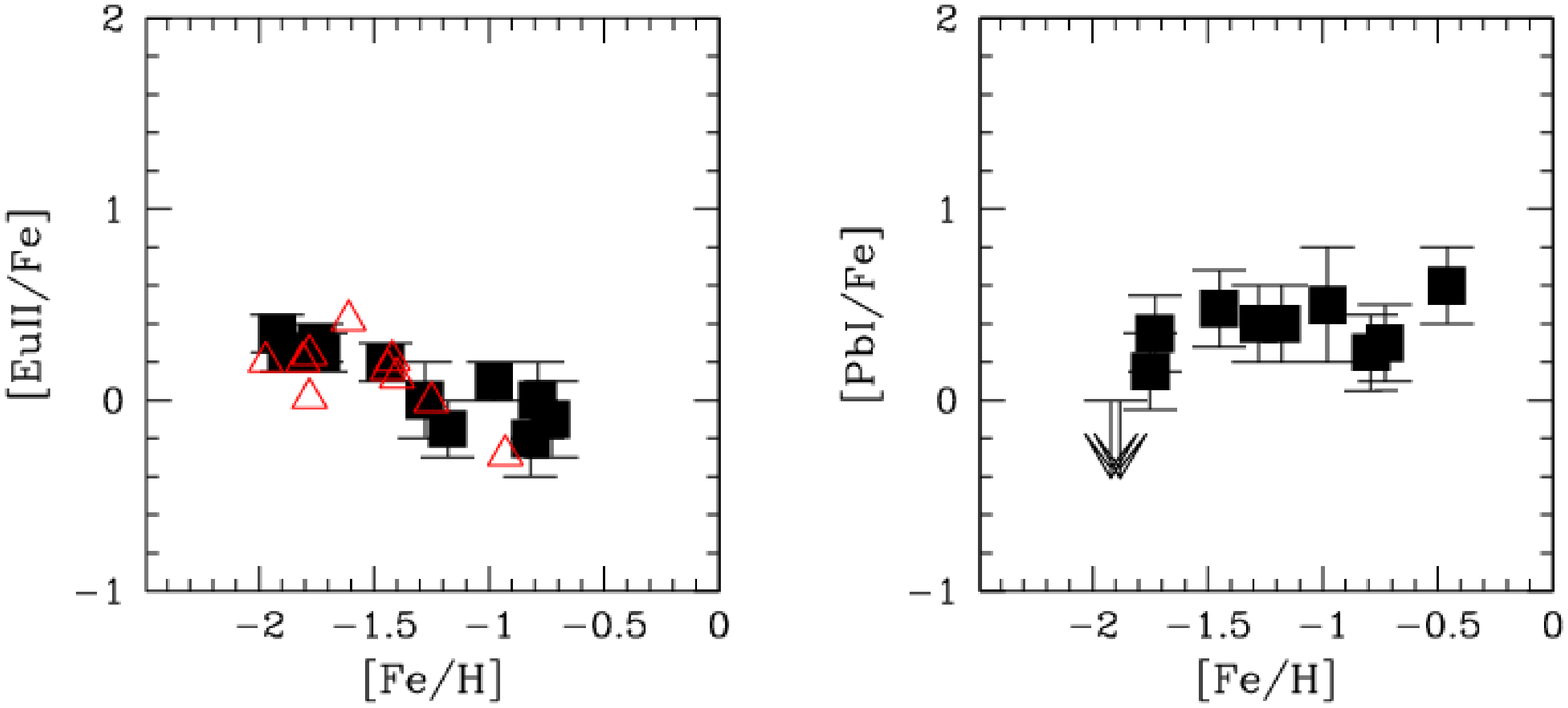}
\caption{Run of [X/Fe] ratios vs [Fe/H] for Y~{\sc ii} and Zr~{\sc ii} (upper panels), 
La~{\sc ii} and Ce~{\sc ii} (middle panels), Pb~{\sc i} and Eu~{\sc ii}
(lower panels). Triangles are data from S00, while 
empty squares are for La estimates from JP10.}\label{f:spro}
\end{center}
\end{figure*}

In Figure~\ref{f:spro} our derived \ncap\ abundances [X/Fe]
are plotted against [Fe/H] metallicity.
For comparison, we also include Y, Zr, and Ce abundances from S00, 
and La abundances from JP10.
With few exceptions the results from the three independent studies
are in agreement. 
In our sample there is a metallicity gap of 0.3~dex at the
low end, with four stars having [Fe/H]~$<$ $-$1.7 and the remaining
eight stars with [Fe/H]~$>$ $-$1.7.
In the discussion below we will contrast these two groups, referring
to the relatively metal-rich ones as those with [Fe/H]~$>$ $-$1.7.

First consider the lighter \ncap\ elements Y and Zr.
We observe a substantial change with metallicity in the 
abundances derived from Y~{\sc ii} and Zr~{\sc ii} lines, as can 
be seen in the upper two panels of Figure~\ref{f:spro}.
Indeed,  for [Fe/H]$\gtrsim$$-$1.7, [Y{\sc ii}/Fe] and [Zr{\sc ii}/Fe] 
appear to suddenly increase by more than a factor of four, reaching
maximum values of 1.0 and 1.5, respectively at the highest 
\ome\ metallicities.
The average value of $<$[Y{\sc ii},Zr{\sc ii} /Fe]$>$ in the four 
most metal-poor stars is $+$0.1, while in stars with [Fe/H]$>-$1.7 
it is $+$0.8, i.e. an increase of of 0.7~dex.
These increases in [Y,Zr/Fe] are far beyond variations that could
be attributed to observational/analytical uncertainties.

Somewhat smaller increases in the relative Y and Zr abundances 
($\sim$0.5 dex) with increasing metallicity were reported by S00.
First, it should be noted that the metallicity ranges of their study 
is somewhat different from ours.
That is, S00 have nine stars out of 10 with [Fe/H]$<-$1.3, and just 
one at [Fe/H]$\sim-$1 dex, while our sample includes seven out of
12 with [Fe/H]$\gtrsim -$1.3 and five more metal-rich than $-$1.0.
This metallicity mismatch makes direct comparison of the variation 
very difficult, and the big changes in Y and Zr become more 
apparent above [Fe/H]~$\sim$ $-$1.5, where our sample has more
representatives.
Additionally, S00 performed an equivalent width (EW) analysis on 
red Y~{\sc ii} and Zr~{\sc ii} lines, which are extremely weak, 
with typical values of only a few m\AA~ and hence may be more affected 
by measurement uncertainties than our stronger transitions.
However, both our sample (12 stars) and the S00 one (10 stars) 
are quite small and there is no evidence in the literature that the 
star-to-star scatter in these elements is small within a 
metallicity bin in \ome. 
A more robust estimation of the trends of these elements with
metallicity awaits an attack on larger samples of cluster members
distributed over the whole \ome\ metallicity range.

Turning to the heavier \ncap\ elements, in the middle left
panel of Figure~\ref{f:spro} we display [La{\sc ii}/Fe] values from this
work along with those from S00 and JP10.
Although the three studies have used different 
instrumental setup(s), spectral features, and abundance codes, 
there is an excellent agreement between our study and the others,
ignoring one very discrepant high-La, low-metallicity star from S00. 
Moreover, as also noted by JP10, our lower La abundances with 
respect to S00 may reflect to a certain extent the lack of the hyperfine 
structure inclusion in that study.
The effect is particularly relevant at higher metallicity, where 
La~{\sc ii} lines become stronger and ``suffer'' the effects of the hyperfine structure to a major degree. 

\begin{figure}[htbp]
\includegraphics[width=8cm]{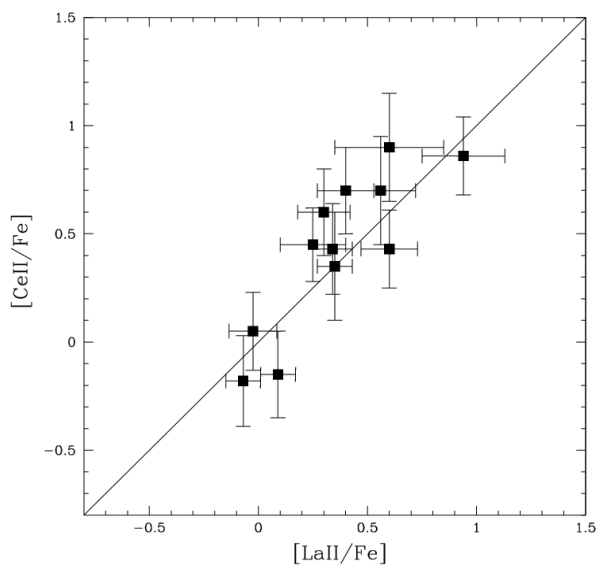}
\caption{[Ce/Fe] as a function of [La/Fe]. The solid line is a 1:1 relation.}\label{f:lace}
\end{figure}

Our Ce abundance trend with \ome\ metallicity (middle panel 
of Figure~\ref{f:spro}) excellently agree with that of La.  
Generally the Ce abundances of S00 agree with ours at similar
metallicities, even though these authors' study had to rely on one
weak Ce~{\sc ii} line at 6052~\AA.
To see the relationship between Ce and La more clearly, we correlate in Figure~\ref{f:lace} [Ce/Fe] with [La/Fe]. 
Clearly, these elements vary in step within observational 
uncertainties, over 1~dex ranges in both metallicity and
relative abundance.
This is a sensible result: in material created by the \spro, 
La and Ce should share a similar abundance trend, and the \spro\
fraction of Ce in the solar system material (i.e. 81\%) is even
higher than La, which has 65\%. 
Note that in Figure~\ref{f:spro} the three most metal-poor stars
may have slightly underabundant Ce abundances.
The \ncap\ elements in lower-metallicity \ome\ stars probably have
been generated in substantial quantities both by the \spro\ and the
\rpro, so a perfect Ce$-$La correlation over the entire metallicity
range should not be expected.

\begin{figure*}[htbp]
\begin{center}
\includegraphics[width=15cm]{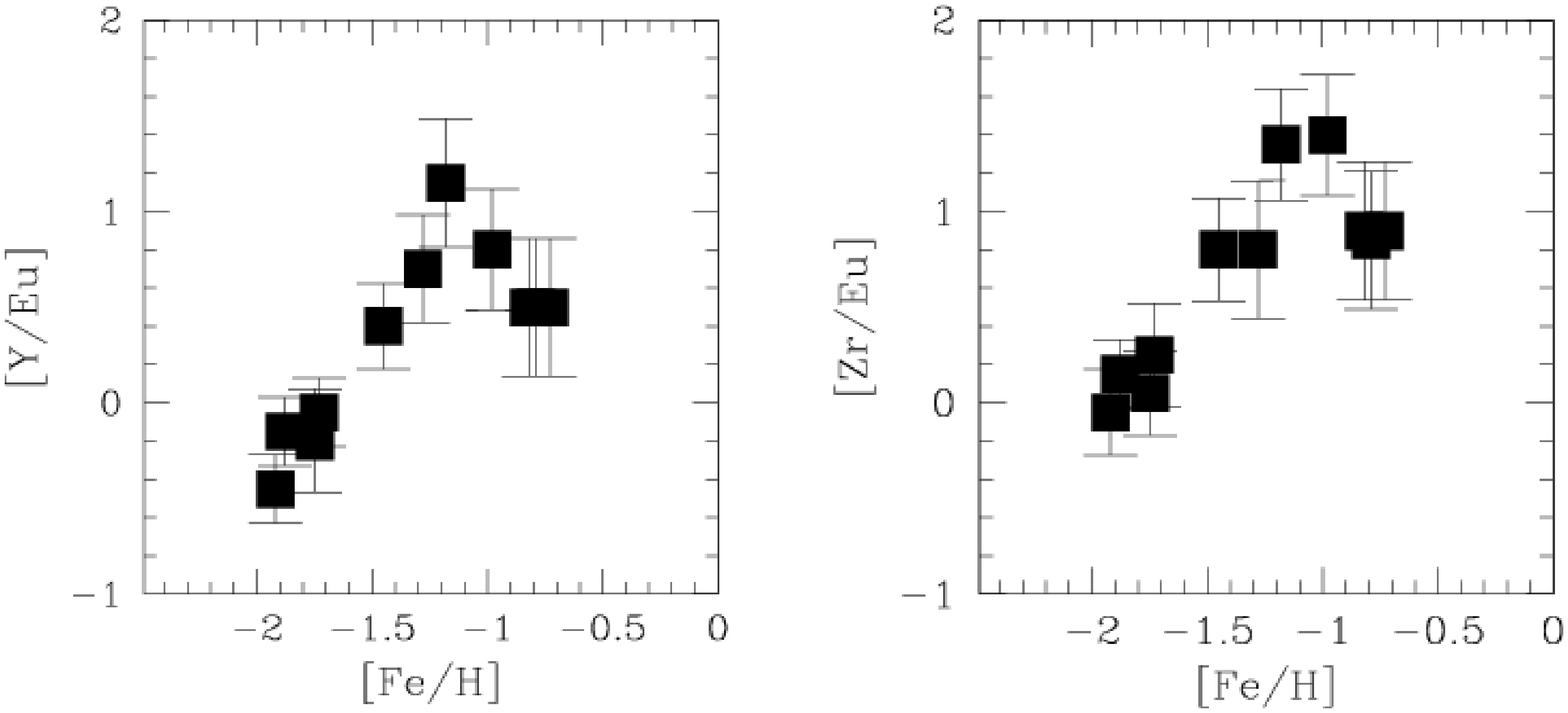}
\includegraphics[width=15cm]{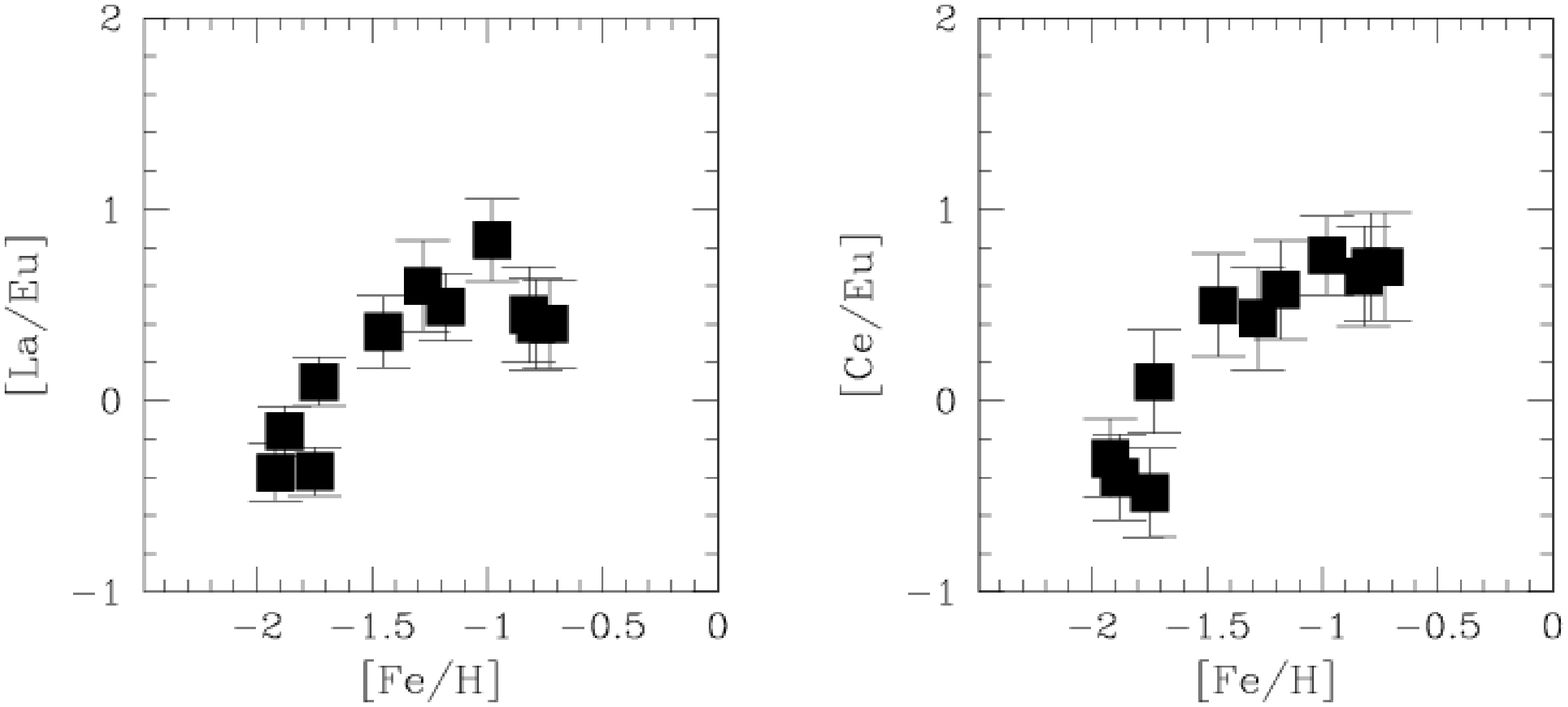}
\includegraphics[width=15cm]{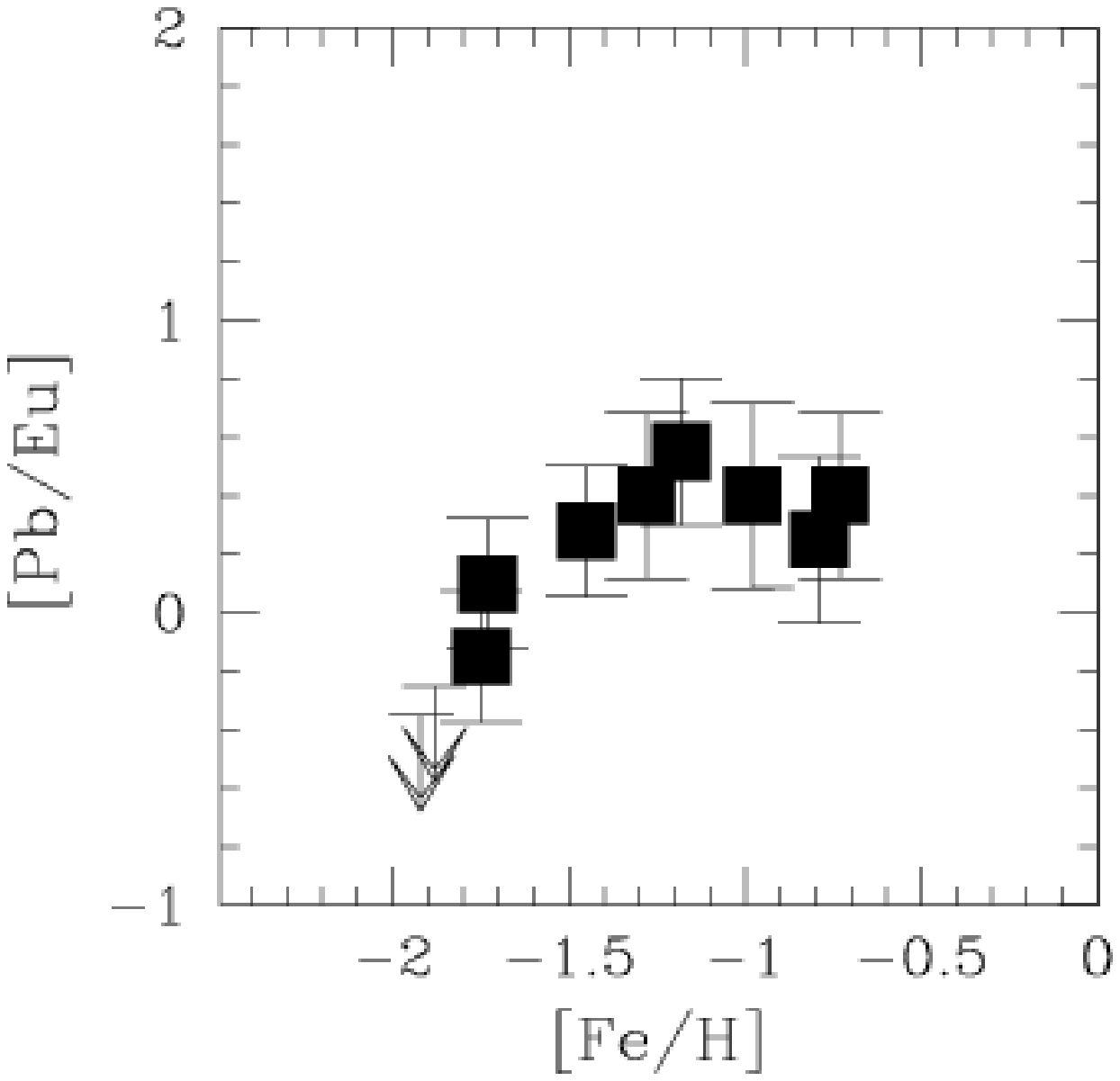}
\caption{Run of [X/Eu] with [Fe/H] for Y, Zr, La, Ce, and Pb.}\label{f:pblaeu}
\end{center}
\end{figure*}

The decreasing contribution of the \rpro\ to the \ncap\ abundances
with increasing metallicity is evident in the bottom left-hand panel of 
Figure~\ref{f:spro}, which displays the run of [Eu~{\sc ii}/Fe]
with [Fe/H].
Eu is the most easily observed \rpro-dominant element,
with a solar-system \spro\ contribution of only 3\% to its
total abundance (e.g. Simmerer et al. 2004). 
The most metal-poor \ome\ stars have [Eu/Fe]~$\approx$ $+$0.3,
a typical value for globular cluster stars.
But [Eu/Fe]~$\sim$ 0.0 at the highest metallicities, a trend
clearly opposite to that of the previously discussed 
\ncap\ elements.\footnote{JP10 determined [Eu/Fe] only for two out of our 12 sample stars, 
namely \#48323, \#34180: analysing the 6645 \AA~feature, they 
obtained [Eu/Fe]=$-$0.08 and +0.27 dex, respectively.
While for star \#48323 their value agrees very well with our 
measurement, i.e. [Eu/Fe]=$-$0.10, we found a lower value for 
star \#34180, namely [Eu/Fe]=0.00. 
A more detailed study of Eu abundances from multiple transitions
in many \ome\ stars should be undertaken.}
To emphasize this change in nucleosyntic dominance, 
in Fig.~\ref{f:pblaeu} we have constructed $s$-/\rpro\ abundance 
surrogates by plotting [X/Eu] ratios 
versus metallicity for Y, Zr, La, Ce, and Pb.
The trends displayed in each panel are qualitatively similar;
these five elements appear to have the same \spro\ nucleosynthetic
origin.

\begin{figure}[htbp]
\includegraphics[width=8cm]{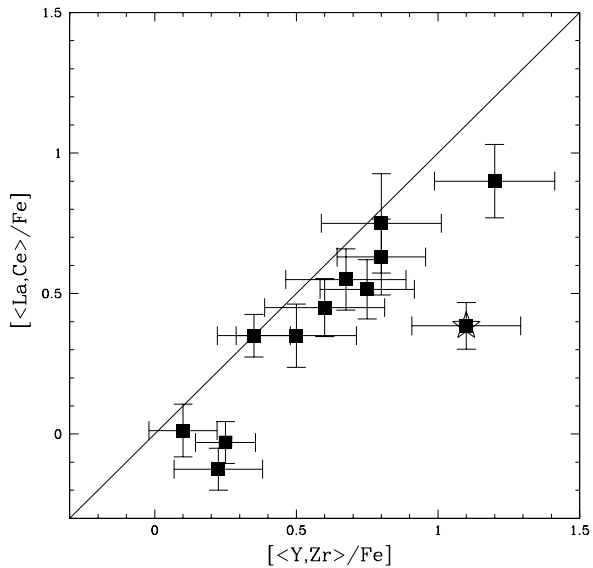}
\caption{Average values for Y~{\sc ii} and Zr~{\sc ii} as a function of 
average La~{\sc ii} and Ce~{\sc ii} abundances.}\label{f:yzrlace}
\end{figure}

The average value of $<$[La~{\sc ii},Ce~{\sc ii}/Fe]$>$ for 
stars with [Fe/H]~$>$ $-$1.7 is 0.57, while the four 
metal-poor stars it is 0.05, i.e. a variation of $\approx$0.5~dex.
As can be clearly seen from Figure~\ref{f:yzrlace}, we found evidence that light-$s$ elements (first-peak) are 
produced more efficiently than the heavy (second-peak) ones. 

S00, on the basis of their data, concluded the opposite; but 
because they used have a narrower metallicity range and
somewhat lower quality data, their conclusion may be affected and our result seem more robust.

The run of [Pb/Fe] as a function of [Fe/H] is shown in 
the bottom right-hand panel of Figure~\ref{f:spro}. 
While in the two most metal-poor stars only an upper limit to the 
Pb abundance can be derived, we detected Pb lines in all stars with 
[Fe/H]$\gtrsim$$-$1.75. 
The distribution resembles those of [La/Fe] and [Ce/Fe] 
shown in the middle panels on Figure~\ref{f:spro}.
Our derived trends of La, Ce, and Pb with metallicity also are in agree with the [La/Fe] and [Ba/Fe] trends shown by JP10 and M11,
i.e, an increase in relative \ncap\ element abundances beginning
at [Fe/H]$\sim$$-$1.7/$-$1.6.

In our \ome\ sample, the ratios [La/Fe], [Ce/Fe], and [Pb/Fe] 
remain roughly flat at metallicity higher than [Fe/H]$\sim$$-$1.5 dex. 
Considering the two upper limits as measurements, we found a linear 
correlation coefficient $r$=0.69, with a significance level $>$99 \%,
comparing the trends of Pb with La, Ce.
Thus the rise of the \spro\ in \ome\ includes elements over the
whole \ncap\ domain, from Y (Z~= 39) at least up to Ce (Z~=58), 
and possibly up to Pb (Z~= 82).
Only the {\it main} component of the \spro\ can yield products
over such a wide atomic number (mass) range.
This argues against dominance by the {\it weak} \spro\ component;
it may have contributed to the overall increase in \ncap\
abundances with metallicity, but the major effect almost certainly
arises from the {\it main} component.

\begin{table*}[htbp]
\caption{Abundances for our sample stars.}\label{t:abu}
\begin{center}
\begin{tabular}{lcccccccccr}
\hline\hline
Leiden   & [Pb~{\sc i}/Fe] & [Y~{\sc ii}/Fe] & [Zr~{\sc ii}/Fe] & [La~{\sc ii}/Fe] & [Ce~{\sc ii}/Fe] & [Eu~{\sc ii}/Fe] & [O/Fe] & [C/Fe] & [N/Fe] & [C+N+O/Fe] \\ 
         &                 &                 &                 &                  &                  &                  &        &  
	      &        &    \\
\hline 
         &                  &                &                 &                  &                  &                  &                &      &                &   \\    
16015 &   $<$0.00  & -0.10$\pm$0.15 & 0.30$\pm$0.20 & -0.03$\pm$0.11 &  0.05$\pm$0.18    &  0.35$\pm$0.10 &  0.39   &  -0.52$\pm$0.10  &  0.42$\pm$0.15  &  0.27  \\		    
37247 &   $<$0.00  &  0.10$\pm$0.15 & 0.40$\pm$0.15 &  0.09$\pm$0.08 & -0.15$\pm$0.20    &  0.25$\pm$0.10 &  0.03   &  -0.87$\pm$0.10  &  1.17$\pm$0.10  &  0.27  \\		    
33011 & 0.15$\pm$0.20 &  0.10$\pm$0.25 & 0.35$\pm$0.20 & -0.07$\pm$0.08 & -0.18$\pm$0.21 &  0.30$\pm$0.10 &  0.40   &  -0.47$\pm$0.10  &  0.32$\pm$0.15  &  0.27  \\		 
41039 & 0.35$\pm$0.20 &  0.20$\pm$0.15 & 0.50$\pm$0.25 &  0.35$\pm$0.08 &  0.35$\pm$0.25 &  0.25$\pm$0.10 &  0.18   &  -0.77$\pm$0.10  &  1.27$\pm$0.10  &  0.39  \\		 
46092 & 0.48$\pm$0.20 &  0.60$\pm$0.20 & 1.00$\pm$0.25 &  0.56$\pm$0.16 &  0.70$\pm$0.25 &  0.20$\pm$0.10 & -0.50   &  -0.77$\pm$0.15  &  $>$1.47	 &  0.40  \\		 
34029 & 0.40$\pm$0.20 &  0.70$\pm$0.20 & 0.80$\pm$0.30 &  0.60$\pm$0.13 &  0.43$\pm$0.18 &  0.00$\pm$0.20 &  0.26   &  -0.37$\pm$0.15  &  1.57$\pm$0.15  &  0.62  \\		 
44462 & 0.40$\pm$0.20 &  1.00$\pm$0.30 & 1.20$\pm$0.25 &  0.34$\pm$0.09 &  0.43$\pm$0.21 & -0.15$\pm$0.15 &  0.70   &	0.43$\pm$0.15  &  0.47$\pm$0.10  &  0.62  \\		 
60066 & 0.50$\pm$0.30 &  0.90$\pm$0.30 & 1.50$\pm$0.30 &  0.94$\pm$0.19 &  0.86$\pm$0.18 &  0.10$\pm$0.10 & -0.27   &  -0.67$\pm$0.15  &  $>$1.47	 &  0.43  \\  	 
60073 &   ...	      &  0.30$\pm$0.30 & 0.70$\pm$0.30 &  0.25$\pm$0.15 &  0.45$\pm$0.17 & -0.20$\pm$0.20 &  0.07   &  -0.47$\pm$0.15  &  $>$1.47	 &  0.49  \\		 
34180 & 0.25$\pm$0.20 &  0.50$\pm$0.30 & 0.85$\pm$0.30 &  0.40$\pm$0.13 &  0.70$\pm$0.20 &  0.00$\pm$0.20 &  0.30   &  -0.47$\pm$0.15  &  $>$1.47	 &  0.56  \\  	 
48323 & 0.30$\pm$0.20 &  0.40$\pm$0.30 & 0.80$\pm$0.30 &  0.30$\pm$0.12 &  0.60$\pm$0.20 & -0.10$\pm$0.20 & -0.22   &  -0.57$\pm$0.15  &  $>$1.47	 &  0.44   \\		 
54022 & 0.60$\pm$0.20 &  0.30$\pm$0.30 & 1.30$\pm$0.30 &  0.60$\pm$0.25 &  0.90$\pm$0.25 &  ...~~~~~~~	  & -0.19   &  -0.57$\pm$0.15  &  $>$1.57	 &  0.52 \\		 
\hline\hline     
\end{tabular}
\end{center}
\end{table*}

\subsection{The C+N+O sum}

\begin{center}
\begin{figure*}[htbp]
\includegraphics[width=15cm]{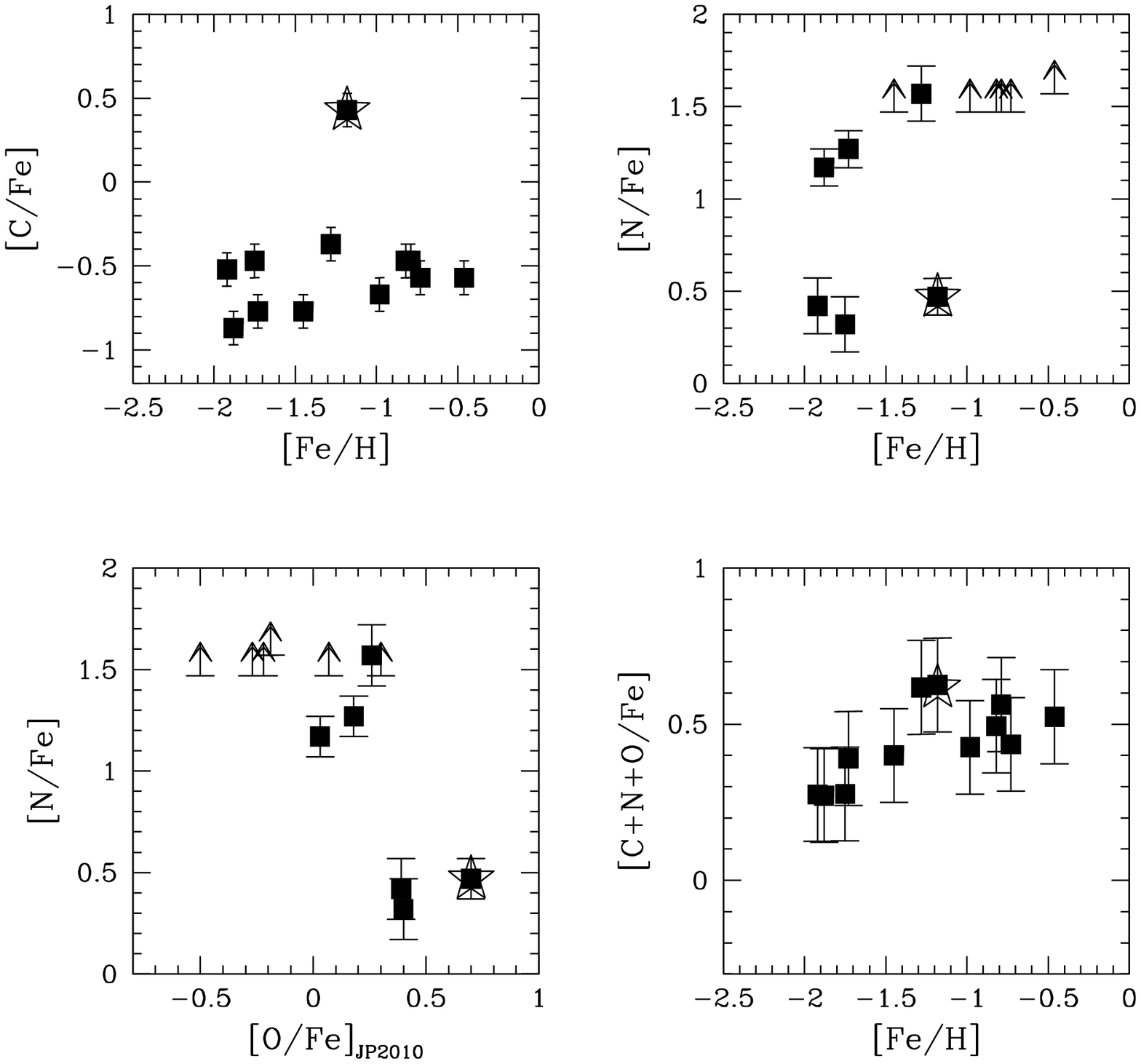}
\includegraphics[width=15cm]{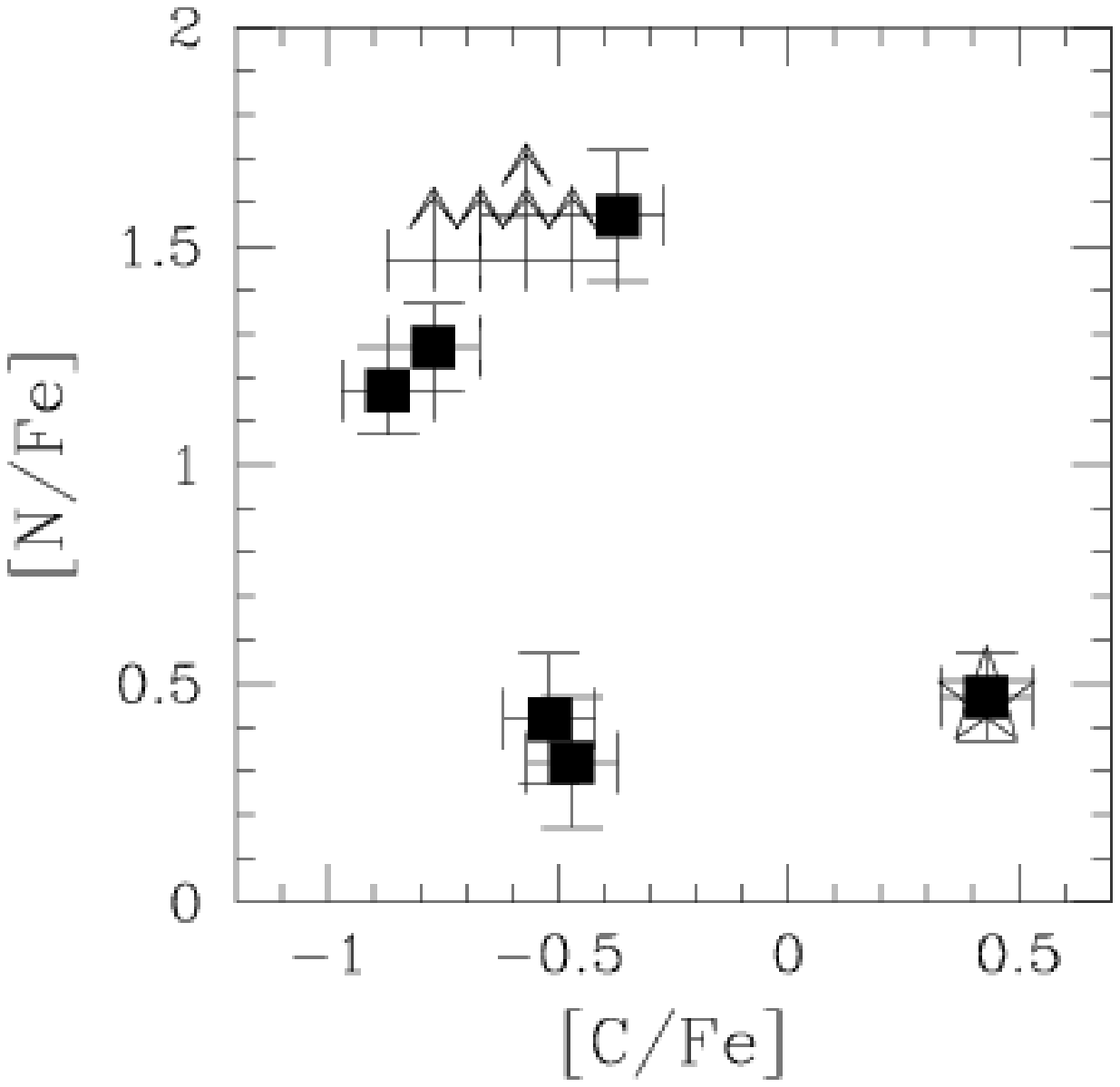}
\caption{Upper panels: [C/Fe] and [N/Fe] vs. the iron content. 
The N-O and C-N planes are shown in the left middle and left lower panels,
respectively.   
Our results on the C+N+O sum is displayed in the right middle panel.
The star
\#44462 is denoted by a star symbol (see text).}\label{f:cno}
\end{figure*}
\end{center}

\begin{figure}[p]
\includegraphics[width=8cm]{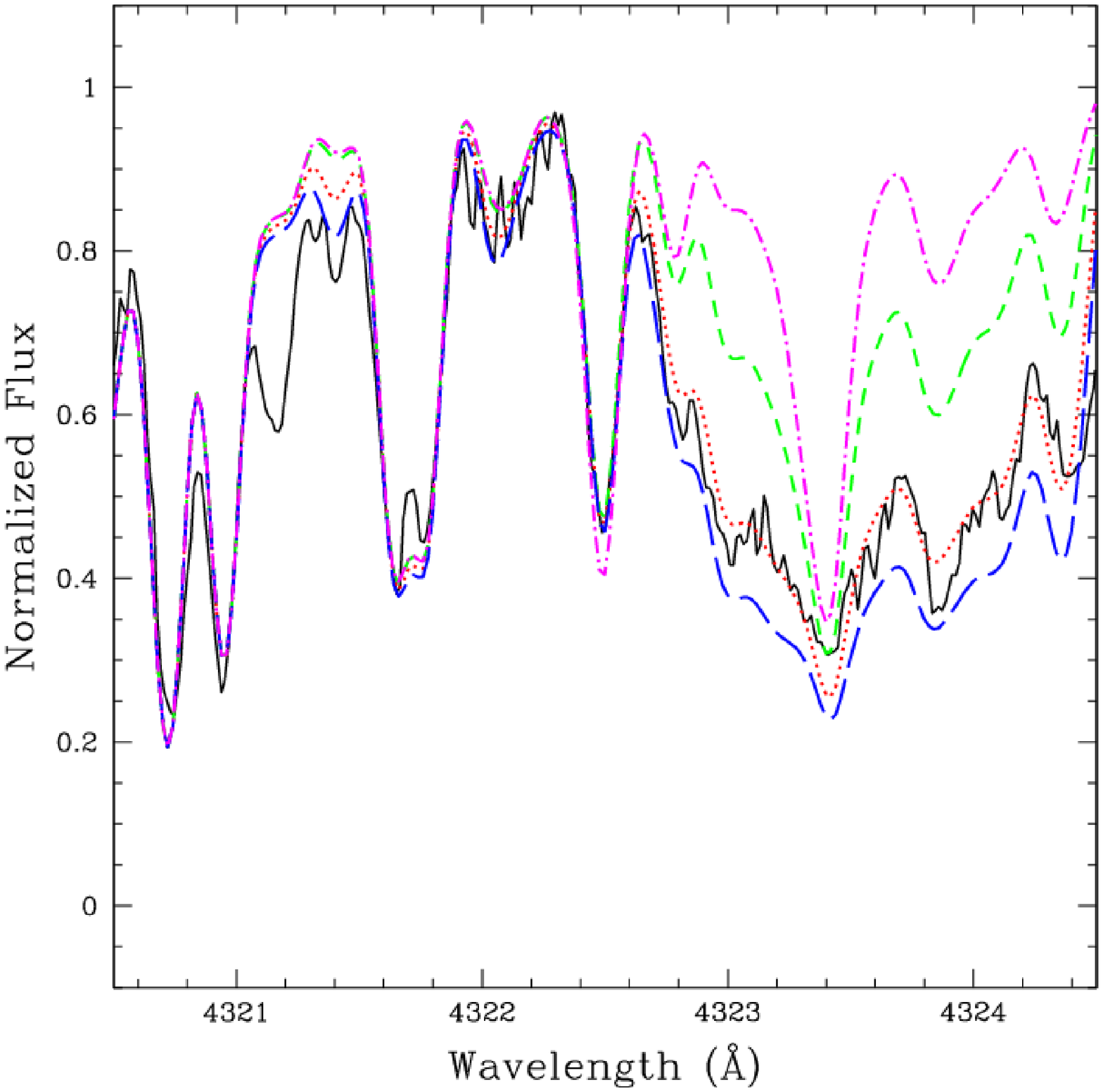}
\caption{Spectral synthesis for the peculiar star \#44462 around the CH G-band at 4300 \AA. 
The continuous line is for the observed spectrum, while dotted lines are for the best fit
at [C/Fe]=+0.4. [C/Fe]=$-$1 (dot-dashed), $-$0.5 (short-dashed), and +0.7 (long-dashed) are also shown.}\label{f:synt}
\end{figure}

In Table~\ref{t:abu} we report our results for C and N abundances 
as derived from the CH and CN bands, and in Fig.~\ref{f:cno} we illustrate
the trends of these elements with other abundance quantities.
For our complete sample we obtained an average value of 
[C/Fe]=$-$0.51$\pm$0.10 (rms=0.33), which agrees very well
with previous estimates (e.g. Norris \& Da Costa 1995; see also JP10 
who adopted that value). 
However, one star, i.e. \#44462, has an anomalously high C abundance.
In all four panels of Fig.~\ref{f:cno} this star is labelled with 
a star symbol, because it clearly has not experienced the same
chemical history as our other \ome\ giants. 
The [C/Fe] ratio for this star is more than 10 times (or 2.5$\sigma$) 
higher than the average value of the other 11 stars. 
The high C content of \#44462 is illustrated in Figure~\ref{f:synt} 
with a small part of the CH G-bandhead. 
Together with the observed spectrum, synthetic spectra with different 
[C/Fe] values ([C/Fe]=$-$1, $-$0.5, +0.4, and +0.7) are plotted. 
The figure demonstrates that [C/Fe]~= $-$0.6 (the average value of 
our complete sample; see below) fails in reproducing the observed CH 
features of \#44462: the best fit for this star is [C/Fe]= $+$0.4.

Star \#44462 does not stand apart from other stars of its
metallicity domain in its total \ncap\ content.
For example, consider stars \#34029 ([Fe/H]~= $-$1.28; 
Table~\ref{t:phot}) and \#60066 ([Fe/H]~= $-$0.98), whose metallicities 
surround \#44462 ([Fe/H]~= $-$1.18). 
From simple means of the abundances given in Table~\ref{t:abu} for 
the \spro\ elements Y, Zr, La, Ce, and Pb, we obtain
[\spro/Fe]~= $+$0.59 for \#34029, $+$0.67 for \#44462, and $+$0.94 for
\#60066.
These mean values appear to vary directly with [Fe/H], as we have
shown above, but are insensitive to [C/Fe].
However, the lighter \ncap\ elements (Y, Zr) are far more overabundant
than the heavier ones (La, Ce) in \#44462 relative to the comparison stars 
(indeed, to any other star of our \ome\ sample).
Therefore whatever process produced the very high [C/Fe] ratio 
in \#44462 apparently yielded no contribution to its
heavier \ncap\ elements.
Detailed nucleosynthetic arguments on the history of this star
are beyond the scope of this paper.

The simplest suggestion for \#44462 is that it could have 
been part of a binary system where the donor star was an AGB, which 
detached only after a few episodes of the third dredge-up.
To search for possible binarity of \#44462, we looked for extant 
radial velocity information.
This star was previously analysed by Mayor et al. (1997) and 
Reijns et al. (2006), who derived v$_{\rm rad}$~=
238.09$\pm$0.27 kms$^{-1}$  and 236.5$\pm$1.1 kms$^{-1}$, respectively. 
From our spectra we found v$_{\rm rad}$=230$\pm$1kms$^{-1}$. 
This is suggestive of radial velocity variability, but
additional spectroscopic  monitoring of this star will be needed to put 
more firm constraints on this issue.

Discarding this ``mild" carbon-star\footnote{
Note that the C/O ratio is not $\gtrsim$1 as the formal definition of a 
carbon star requires.}, 
we obtained an average of [C/Fe]=$-$0.59$\pm$0.04 (rms=0.16). 
Our estimate excellently agree with the value derived for metal-poor 
($-$2$<$[Fe/H]$<-$1) field giants studied by Gratton et al. (2000): 
the average carbon-to-iron abundance for stars on the upper RGB
(i.e. more luminous than the RGB-bump) is [C/Fe]=$-$0.58$\pm$0.03 (rms=0.12).

Once a C abundance was determined for each star, we adopted 
its value to derive the CN product from the strong violet band at 4215~\AA.
As can be seen in Fig.~\ref{f:cno} (right upper panel), we found 
that while low-metallicity stars can be either N-rich or N-poor stars, 
the most metal-rich ones are all N-rich. 
For stars with [Fe/H]~$\gtrsim$ $-$1.4 (with a sharp dependence also 
on temperature), only lower limits can be provided for nitrogen 
abundances, because the CN bandhead becomes strongly saturated. 
Given that our spectra did not include the forbidden [O~{\sc i}] line at 
6300~\AA~ (or the permitted triplet, which is
too weak in giants however) we could not derive oxygen abundances. 
Therefore, we adopted O values from JP10. 
On the basis of our N measurements we confirm that a clear N$-$O 
anticorrelation is observable for \ome~giants, with the exception of the most metal-rich stars 
(see also M11, JP10).
The C+N+O sum as a function of [Fe/H] is shown in the bottom panel 
(right-hand side) of Figure~\ref{f:cno}. 
We see a moderate rise of this sum up to [Fe/H]$\sim-$1.5, and then the
C+N+O values remain constant at higher metallicity.
This again recalls the imprinting of low-mass AGB contribution 
(see e.g., Busso et al. 1999 and references therein) to intra-cluster
pollution, along with the more massive stars that usually are responsible 
for internal chemical enrichment in GCs. 
\section{Discussion}\label{s:disc}

As discussed in the previous sections, we found that the ratio of 
light $s$-process element (Y, Zr) over the heavy $s$-process ones 
(La, Ce) is notably shifted towards the first one. 
However, to obtain a robust evidence of enrichment mechanisms and 
timescales and to discuss on more quantitative grounds the 
implications of our results, the computation of the [$hs$/$ls$]\footnote{
$hs$ is for heavy $s$-process elements (second-peak, here La and Ce), and 
$ls$ is for the lighter ones (first-peak, here Y and Zr). 
Lead (third-peak) is considered separately.} 
ratio is needed. 
That is, for all the neutron-caption elements, the $r$-fraction
contribution has to be removed to allow insights into the $s$-process
nucleosynthesis and the involved mass range, and the contamination 
from different production site(s) has to be derived.

To do this, we adopted for Y, Zr, La, Ce, Eu, the $s$- and $r$-fractions 
listed in Sneden \& Parthasarathy~\citep{sn83},  
while for Pb we chose values given by Plez et al. (2004, see Table~\ref{t:fractions}).
\begin{table}
\begin{center}
\caption{Adopted $r$/$s$ fractions for $n$-capture elements}\label{t:fractions}
\begin{tabular}{lccr}
\hline\hline
 Specie   &  $s$-fraction & $r$-fraction & Reference$^{*}$ \\
          &      \%         &    \%          &            \\
          &                &             &                \\
\hline	  
 Y        &  72.9          &  27.1       & [1] \\
 Zr       &   79.1	   &  20.9	 & [1] \\
 La       &   64.9	   &  35.1	 & [1] \\
 Ce       &   80.6	   &  19.4	 & [1] \\
 Eu       &    8.5	   &  91.5 	 & [1] \\
 Pb       &   98.0	   &   2.0  	 & [2] \\
 \hline
 \end{tabular}
 \begin{list}{}{}
 \begin{footnotesize}
 \item[$^\mathrm{*}$] [1] Sneden \& Parthasarathy (1983); [2] 
 Plez et al. (2004)
\end{footnotesize}
\end{list}
\end{center}
\end{table}
We proceeded as follows. 
First, we assumed that 
[Eu$_s$/La$_s$]$\simeq{[{\rm Eu}_s/{\rm La}_s]}_{\odot}$ for 
all our sample stars, i.e.  the ratio of $s$-components of Eu and 
La is solar, because these elements are close to each other
in the neutron-capture chain. 
This is a reasonable assumption because the same $s$-nucleosynthesis 
for elements with similar neutron numbers should be expected.
We then derived for each element of each star 
the $s$-process contribution to the total abundance.
The first consideration is that, as also noted by several authors 
(e.g., Truran 1981; S00; JP10; M11), the most metal-poor stars have a composition 
dominated by the $r$-process. 
This means that even when we deal with species that in the Sun 
are mainly produced by the $s$-process (e.g., La, which in the solar 
system is 35\% $r$-process and 65\% $s$-process), the heavy-element 
nucleosynthesis is mostly $r$-process in metal-poor stars. 
As an example, the metal-poor star \#16015 ([Fe/H]~= $-$1.92) has
an A(La)~= $-$0.820 ([La/Fe]~= $-$0.03, see Table~\ref{t:abu}), of which only 
$\sim$10\% should come from s-process production, i.e. A(La)$_s$=$-$1.838. 
Thus, to match the condition 
[Eu$_s$/La$_s$]~$\simeq$ [${\rm Eu}_s/{\rm La}_s]_{\odot}$, 
when the nucleosynthesis is markedly $r$-process type, these 
stars cannot be considered in the [$hs$/$ls$] plane, because their 
ratios are not genuine tracers of elements produced by AGB stars 
and even small errors in abundances cause large uncertainties in 
the estimated $s$-fraction.

\begin{figure}[htbp]
\includegraphics[width=8cm]{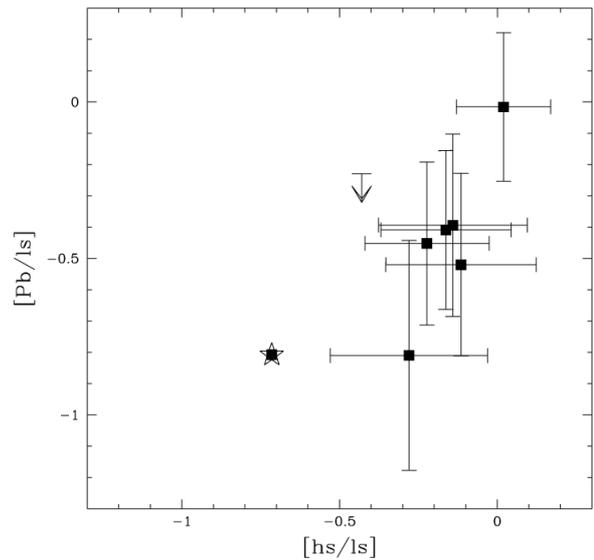}
\caption{[Pb/$ls$] as a function of the [$hs$/$ls$] ratio (see text for discussion).}\label{f:hsls}
\end{figure}

In Figure~\ref{f:hsls} we show the [Pb/$ls$] ratio as a function 
of [$hs$/$ls$] for all our sample stars whose neutron-capture
element pattern is predominantly $s$-process.
Excluding the anomalous C-rich star \#44462 discussed above, 
there is a very clear positive correlation in the 
[Pb/$ls$] $vs$ [$hs$/$ls$] diagram. 
As expected, the variation of [Pb/$ls$] is roughly a factor of two 
larger than the one [$hs$/$ls$]: this simply reflects the proximity 
of first- and second-peak elements. 
The Pb variation evident in Figure~\ref{f:hsls} appears to 
rule out dominant contributions from the {\it weak} component 
to generation of the total $s$-process pattern observed in
\ome: the main component clearly is at work here.
The suggestion of constant [Cu/Fe] as a function of [Fe/H] 
(e.g., Cunha et al. 2002) also converges towards the same 
conclusion. 
Moreover, theoretical studies (e.g. Raiteri et al. 1992; 
Pignatari \& Gallino 2008) argue that the {\it weak} component 
cannot produce a significant amount of elements heavier than Y,Zr at low
metallicity (see references therein for details). 

\begin{figure}[htbp]
\begin{center}
\includegraphics[width=8cm]{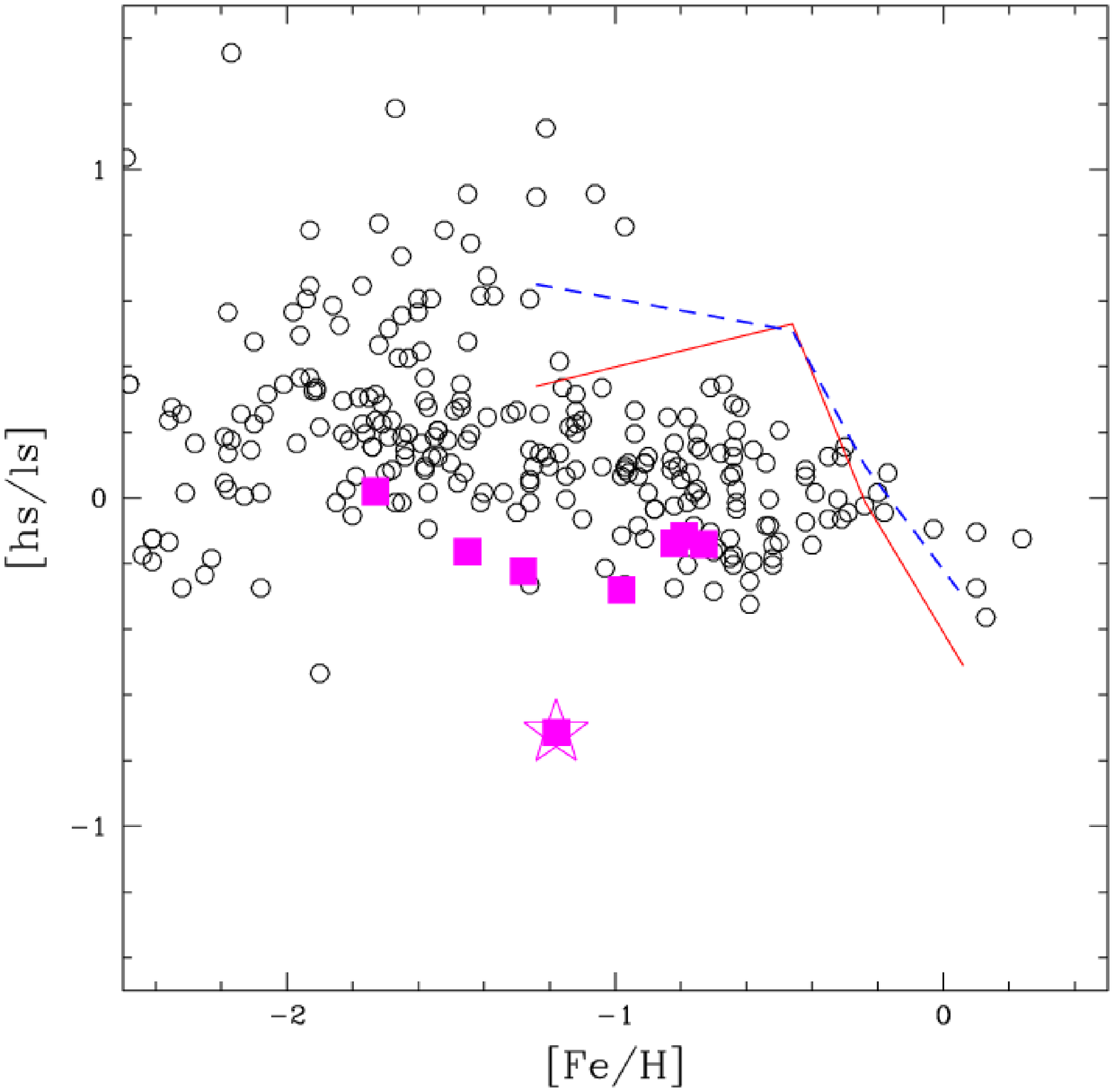}
\caption{Run of the [$hs$/$ls$] ratio as a function of [Fe/H] for field stars (both thick disk and halo, Venn et al. 2004) 
and \ome~giants (filled symbols). The solid and dashed curves are for 3 M$_\odot$ AGB and 1.5 M$_\odot$ AGB
models from Cristallo et al. (2009), respectively.}\label{f:field}
\end{center}
\end{figure}

Most tantalising is that, because the variation with metallicity of
$ls$ elements is larger than the $hs$ ones, the neutron exposures 
should be quite small; this implies few thermal pulses and larger masses. 
The $s$-process seen in \ome~ is clearly different from the standard 
``Galactic-like" {\it main} component, which is predicted to arise
from a mass range of donor stars $\approx$1$-$3.5 M$_\odot$.
These AGB stars should yield an over-production of $hs$ with respect to 
$ls$ in stars with metallicities similar to the \ome\ ones. 
In Figure~\ref{f:field} the run of [$hs$/$ls$] ratio with metallicity 
is shown for field stars (open squares, Ba for [$hs$] and Y for [$ls$] 
are from Venn et al. 2004 and references therein) and \ome: as is clear 
from the plot, the [$hs$/$ls$] values for cluster giants are lower 
than in field stars at the same metallicity, indicating an over-abundance 
of light $s$-process elements with respect to the heavier ones.
For a qualitative comparison, because systematic offsets between 
theoretical yields and observations can be present, we also show 
the 3M$_\odot$ (solid curve) and 1.5M$_\odot$ (dashed curve) AGB 
models by Cristallo et al. (2009\footnote{
Available at http://fruity.oa-teramo.inaf.it:8080/modelli.pl}). 
We note that, discarding the peculiar carbon star \#44462, 
the increasing [$hs$/$ls$] ratios occurring at [Fe/H]$\approx-$1 dex 
are well reproduced by the 3M$_\odot$ AGB model, while the 1.5M$_\odot$ 
curve exhibits an opposite (declining) run. 
Additionally, the expected difference between the two models 
($\sim$0.2 dex) agrees well with the average difference 
among \ome\ and field stars.

We therefore think that our data reveal the fingerprint of a 
``peculiar" {\it main} component, biased towards the upper end 
of the mass range of stars involved in the {\it main} $s$-process. 
The observational clue achieved from the $s$-process elements is 
further confirmed by the C+N+O sum, which only slightly increases 
with metallicity, indicating that only stars with a limited number 
of third dredge-up episodes contributed to intra-cluster pollution.

While a quantitative estimate of the mass range requires modelling 
of AGB stars, we tentatively identify stars with mass 
$\gtrsim$3M$_{\odot}$ as responsible for the $s$-process production.  
We do not need to invoke an age spread of one Gyr or more 
(as would be required from the evolution of $\approx$1.5 M$_\odot$ AGBs) 
for the later generations in \ome. 
Because $\approx$3$-$4M$_\odot$ AGB stars do evolve in several 
hundred Myr, a drastic reduction in the age difference of the 
various stellar populations in \ome~ is then suggested.
Recently, D'Antona et al. (2011) concluded that the age spread in 
\ome\ populations can be at most a few $\sim10^8$ yr 
(see also Sollima et al. 2005b), but to reconcile their 
predictions with the observed $s$-process pattern, an unknown site 
of $s$-process element production has to be invoked. 
D'Antona et al. advanced the hypothesis that carbon-burning shells 
of the low-mass tail of the SN II progenitors could be that 
production site, referring to the study by The et al. (2007). 
However, these authors pointed out that massive stars (at the final 
stages of their lives) contribute at least for 40\% to $s$-nuclei 
with mass A$\leq$87, but only for $\sim$7\% (on average) for 
heavier nuclei (i.e. A$>$90). 
The detected Pb variation hence seems to contradict this scenario, 
suggesting again that the {\it main} \spro\ component must be a significant mechanism in producing the \ome\ \ncap\ element mix. 

We note that our results do not critically depend on precise 
values of the $r$/$s$ fractions adopted in the computation 
of the [$hs$/$ls$] ratios. 
For instance, we used for the Pb $s$-fraction the value of 98\% provided 
by Plez et al. (2004), who stressed, however, that the fraction of Pb 
produced by the $r$-process in the solar system is practically unknown.
Had we instead adopted the 81\% given by Simmerer 
et al. (2004), our results would not have suffered a 
significant change.

In the derivation of [$hs$/$ls$] ratios we used the second-peak 
elements La and Eu as proxies for the $s$ and $r$ fractions.
We checked also if the $r$/$s$ fractions of the \ncap\ elements
could be confirmed with just a sensitive first-peak element ratio, 
e.g. [Rb/Zr].
Rb is a good \rpro\ indicator, because it has a substantial \rpro\ 
component in solar-system material.
Rb~{\sc i} has only one transition at 7800\AA, thus we have the correct 
spectral coverage only for two of the most metal-rich stars, 
namely \#60066 and \#60073. 
However, owing to the low S/N, it was not possible to derive a reliable 
[Rb/Fe] for star \#60066, so we just inspected the trend of \#60073.
As previously done for Eu and La, we assumed that also 
[Rb$_s$/Zr$_s$]~$\simeq$ $[{\rm Rb}_s/{\rm Zr}_s]_{\odot}$: for
this star we obtained the same fraction of \spro\ and \rpro\ 
for each element as the one obtained from the second-peak element ratios.

In the final remarks of their Section 4.4, S00 compared the [Rb/Zr] 
ratio (as a tracer of the neutron density environment, see their 
work for details) with AGB models of 1.5 M$_\odot$, 3 M$_\odot$, 
and 5 M$_\odot$, with different masses of the $^{13}$C pocket
(Gallino et al. 1998, Busso et al. 1999). 
Their plots show that all 5 M$_\odot$ AGB models fail in reproducing 
the observed abundance pattern: this means that the 
$^{22}$Ne($\alpha$,n)$^{25}$Mg reaction, activated in AGB with masses 
$\gtrsim$5M$_{\odot}$, cannot be invoked as responsible for the \spro\
production.
This idea has been recently advanced also by M11. 
In the attempt to reconcile timescale problems, M11 proposed either 
an \spro\ production through the {\it weak} component, or 
intermediate-mass AGBs providing neutrons through $^{22}$Ne 
(see also D'Antona et al. 2011). 
If stars with mass $\gtrsim$5M$_{\odot}$ were neutron-capture element 
producers, then a quite close relationship should exist between \spro\ 
elements and hot H-burning products (e.g., N). 
Owing to the smallness of our sample, we cannot draw significant 
conclusions from our data (recall that at [Fe/H]$>-$1.5, all our sample 
stars are N-rich), but we refer the reader to Gratton et al. (2011) for 
a discussion on the complex run {\bf of} [O/Na] with [La/Fe], based on the 
extensive dataset from JP10.

Our findings instead agree better with the S00's view that 
variations in $s$-process elements in \ome~ are caused by lower-mass AGBs.
Those authors  concluded indeed that the best fit to their observations 
was provided by the lower masses, i.e. 1.5 M$_\odot$, while our 
conclusions converge towards higher masses.
However, we caution the reader again that from their Figure 14 only 
the most metal-poor stars, which should have a more markedly \rpro\ 
pattern  however, can be satisfactorily reproduced by the lower-mass AGB model. 
For stars with [Fe/H] $\geq$$-$1 (where the $s$-process is predominant) 
the 3M$_\odot$ model seems to provide a better agreement. 
Unfortunately, because we could derive the Rb abundances for only 
one star, we cannot build up a similar diagram to that given by S00.
For star \#60073 we found [Rb/Zr]~= $-$0.65 ([Rb/Fe]~= 0.05$\pm$0.05), 
which agrees well with the 3M$_\odot$ AGB model. 
Remember that the comparison of abundances $vs$ AGB models cannot be
conclusive, because theoretical models strongly depend on a free 
parameter, i.e.  the $^{13}$C pocket mass. 
However, although on the basis of only one star, we suggest that 
intermediate-mass AGB stars (5-8 M$_\odot$) would result in an 
overproduction of Rb with respect to Zr (Garcia-Hernandez et al. 2006, 
and references therein), which is not observed.
Once again, this evidence agrees with the previous ones: 
relatively low-mass AGBs appear to have been responsible for the \spro\ 
element enrichment in \ome.

\section{Conclusions}
We presented abundances for a sample of 12 red giants in 
the peculiar globular cluster \ome.  
We derived light $s$-process elements (Y, Zr), heavy $s$-process 
elements (La, Ce), the $r$-process element Eu, and for the first time 
we measured Pb abundances in this cluster. 
As a complementary information we also obtained abundances for C and N 
and we computed the C+N+O sum (retrieving O values from JP10).

We detected an indication for Pb production, occurring at 
[Fe/H]$\gtrsim$$-$1.6: this result allows us to discard the 
{\it weak} component from massive stars as the dominant mechanism
for \ncap\ production in \ome.  
Moreover, from computed [$hs$/$ls$] ratios, we conclude that 
light \spro\ elements (Y, Zr) vary more than the second-peak ones (La, Ce).

On more general grounds, we suggest that the {\it main} \spro\
component active in \ome~ tends towards the higher mass, 
$\gtrsim$3.0 M$_{\odot}$, notably reducing enrichment timescales from 
several Gyr, as needed in the case of 1.2$-$1.5 M$_\odot$ AGBs, to only  
hundreds million years. 

Finally, we note that the acquisition of a more comprehensive sample of 
\ome~stars is of paramount importance to draw definite conclusions 
on these issues.
The difference in abundances between our study and the one by S00 
might not be entirely caused to the analysis, and a real scatter at 
any metallicity bin could be present. 
Accurate heavy-elements abundance estimates for a larger number of
stars could allow us to understand both the run of the \spro\ 
elements with iron and the presence (if any) of an internal variation.

\begin{acknowledgements}
This publication made extensive use of the SIMBAD database, operating
at CDS (Strasbourg, France) and of NASA's Astrophysical Data System. 
This work was partially funded by the Italian MIUR under PRIN
20075TP5K9 and by PRIN INAF ``Formation and Early evolution of Massive star
clusters". 
We thank Marco Pignatari for illuminating discussions on the AGB 
nucleosynthesis. Financial support to CS from the U.S. National Science 
Foundation through grant AST-0908978 is gratefully acknowledged.
We thank the anonymous referee for her/his very helpful suggestions and comments.
\end{acknowledgements}


\begin{thebibliography}{}

\bibitem[1997]{an97}  
Anderson, J. 1998, PhD Thesis, Univ. California, Berkeley

\bibitem[]{aok02}     
Aoki, W., Ryan, S.G., Norris, J.E., et al. 2002, ApJ, 580, 1149

\bibitem[]{armosky}
Armosky, B.J., Sneden, C., Langer, G.E., \& Kraft, R. P. 1994, AJ, 108, 1364

\bibitem[]{barbuy}
Barbuy, B., Zoccali, M., Ortolani, S., et al. 2009, A\&A, 507, 405

\bibitem[2004]{b04}   
Bedin, L., Piotto, G., Anderson, J., et al. 2004, ApJ, 605, L125

\bibitem[2008]{b08}   
Bellazzini, M., Ibata, R.A., Chapman, S.C., et al. 2008, AJ, 136, 1174 

\bibitem[2010]{be10}  
Bellini, A., Bedin, L.R., Piotto, G., et al. 2010, ApJ, 140 631 

\bibitem[2000]{bie00} 
Bi{\'e}mont, E., Garnir, H.~P., Palmeri, P., Li, Z.~S., 
\& Svanberg, S.\ 2000, MNRAS, 312, 116

\bibitem[2011]{bra11}
Bragaglia, A., Sneden, C., Carretta, E., Gratton, R.G., Lucatello, S. 2011, ApJ, submitted

\bibitem[2000]{bu00}  
Busso, M., Gallino, R., \& Wasserburg, G. J. 1999, ARA\&A, 37, 239

\bibitem[2009]{c09a}  
Carretta, E., Bragaglia, A., Gratton, R.G., et al. 2009a, A\&A, 505, 117

\bibitem[2009]{c09b}  
Carretta, E., Bragaglia, A., Gratton, R.G., \& Lucatello, S. 2009b, 
A\&A, 505, 139

\bibitem[2010]{c10a}  
Carretta, E., Gratton, R.G., Lucatello, S., et al. 2010b, ApJ, 722, L1 

\bibitem[2010]{c10b}  
Carretta, E., Bragaglia, A., Gratton, R.G., et al. 2010c, A\&A, 520, 95

\bibitem[2011]{c11}
Carretta, E., Lucatello, S., Gratton, R.G., Bragaglia, A., D'Orazi, V., 2011, A\&A, in press, arXiv: 1106.3174

\bibitem[1997]{cat97} 
Catelan, M. 1997, ApJ, 478, L99

\bibitem[2011]{chiappini}
Chiappini, C., Frischknecht, U., Meynet, G., et al. 2011, Nature, 474, 666  

\bibitem[1999]{coh99}
Cohen, J.G. 1999, AJ, 117, 2434

\bibitem[1986]{cs86} 
Cottrell, P.~L., \& Sneden, C.\ 1986, A\&A, 161, 314

\bibitem[2009]{cr09}  
Cristallo, S., Straniero, O., Gallino, R., et al. 2009, ApJ, 696, 797

\bibitem[2002]{cun02} 
Cunha, K., Smith, V.V., Suntzeff, N.B., et al. 2002, AJ, 124, 379 

\bibitem[2007]{dec}   
Decressin, T., Meynet, G., Charbonnel, C., Prantzos, N., \& Ekstr\"{o}m, 
S. 2007, A\&A, 464, 1029 

\bibitem[2011]{d01}   
D'Antona, F., D'Ercole, A., Marino, A.F., et al. 2011, ApJ, in press, 
arXiv:1105.0366


\bibitem[2009]{dmink} 
de Mink, S.E., Pols, O.R., Langer, N., \& Izzard, R.G. 2009, A\&A, 507, L1

\bibitem[1989]{denis}
Denisenkov, P.A., Denisenkova, S.N. 1989, A.Tsir., 1538, 11

\bibitem[2009]{ds09}  
de Silva, G.M., Gibson, B.K., Lattanzio, J., \& Asplund, M. 2009, 500, 25

\bibitem[2010]{d10}   
D'Orazi, V., Gratton, R.G., Lucatello, S., et al. 2010, ApJ, 719, L213

\bibitem[2004]{fer04} 
Ferraro, F.R., Sollima, A., Pancino, E., et al. 2004, ApJ, 603, L81

\bibitem[2007]{ful07} 
Fulbright, J.P., McWilliam, A., \& Rich, R.M. 2007, ApJ, 66, 1152 

\bibitem[1998]{g98}   
Gallino, R., Arlandini, C., Busso, M., et al. 1998, ApJ, 497, 388  

\bibitem[2006]{gar06} 
Garcia-Hernandez, D.A., Garcia-Lario, P., Plez, F., et al. 
2006, Science, 314, 1751

\bibitem[1985]{gr85}
Gratton, R.G. 1985, A\&A, 148, 105

\bibitem[1988]{gr88}  
Gratton, R.G. 1988, Rome Observatory Preprint Ser., 29  

\bibitem[2000]{gr00}  
Gratton, R.G., Sneden, C., Carretta, E., \& Bragaglia, A. 2000, A\&A, 354, 169


\bibitem[2004]{gr04}  
Gratton, R.G., Sneden, C., \& Carretta, E. 2004 ARA\&A, 42, 385

\bibitem[2010]{gr10}  
Gratton, R.G., Carretta, E., Bragaglia, A., Lucatello, S., D'Orazi, V. 
2010, A\&A, 517, 81 

\bibitem[2011]{gr11}
Gratton, R.G., Johnson, C.I., Lucatello, S., D'Orazi, V., Pilachowski, C. 2011, A\&A, in press, arXiv:1105.5544

\bibitem[2009]{han}

Han, S.-I, Lee, Y.-W, Joo, S.-J., et al. 2009, ApJ, 707, L190

\bibitem[1982]{han82}
Hannaford, P., Lowe, R.M., Grevesse, N., et al. 1982 Astrophys. J. 261, 736

\bibitem[2004]{james}
James, G., Fran\c{c}ois, P., Bonifacio, P., et al. 2004, A\&A, 427, 825

\bibitem[2010]{jp10}  
Johnson, C.I., \& Pilachowski, C.A. 2010, ApJ, 722, 1373 (JP10)

\bibitem[2008]{kay08} 
Kayser, A., Hilker, M., Grebel, E.K., \& Willemsen, P.G. 2008, A\&A, 486, 437

\bibitem[1994]{kr94}  
Kraft, R.P. 1994, PASP, 106, 553

\bibitem[1985]{kur85}
Kurucz, R.L., Furenlid, I., Brault, J., \& Testerman, L. 1985, S\&T, 70, 38

\bibitem[1993]{kur93} 
Kurucz, R.L. 1993, CD-ROM 13, Cambridge, Mass., Smithsonian 
Astrophysical Observatory 

\bibitem[1981]{lr81}  
Lambert, D.L., \& Ries, L.M. 1981, ApJ, 248, 228

\bibitem[1993]{langer}
Langer, G.E., Hoffman, R., \& Sneden, C. 1993, PASP, 105, 301

\bibitem[2001]{l01a}  
Lawler, J. E., Bonvallet, G., \& Sneden, C. 2001a, ApJ, 556, L452

\bibitem[2009]{la09}   
Lawler, J. E., Sneden, C., Cowan, J.J., Ivans, I.I., 
\& Den Hartog, E.A. 2009, ApJS, 182, L51 

\bibitem[2001]{l01b}   
Lawler, J. E., Wickliffe, M. E., den Hartog, E. A., \& Sneden, C. 
2001b, ApJ, 563, 1075

\bibitem[1999]{l99}   
Lee, T.-W., Joo, J.-M., Sohn, Y.-J., et al. 1999, Nature, 402, 55

\bibitem[2009]{li09}   
Lind, K., Primas, F., Charbonnel, C., Grundahl, F., \& Asplund, M. 
2009, A\&A, 503, 545

\bibitem[2006]{lju06}
Ljung, G., Nilsson, H., Asplund, M. \& Johansson, S. 2006, A\&A, 456, 1181.

\bibitem[2009]{mar09} 
Marino, A.F., Milone, A. P., Piotto, G., et al. 2009, A\&A, 505, 1099

\bibitem[2011]{mar11} 
Marino, A.F., Milone, A. P., Piotto, G., et al. 2011, ApJ, 731, 64 (M11)

\bibitem[2011]{martell}
Martell, S.L. 2011, AN, 332, 467

\bibitem[2007]{masho} 
Mashonkina, L.I., Vinogradova, A.B., Ptitsyn, D.A., Khokhlova, V.S., 
\& Chernetsova, T.A. 2007, ARep, 51, 903 

\bibitem[1997]{may97} 
Mayor, M., Meylan, G., Udry, S., et al. 1997, AJ, 114, 1087 

\bibitem[2008]{mil08} 
Milone, A.P., Bedin, L., Piotto, G., et al. 2008, ApJ, 673, 241 

\bibitem[1995]{nd95}  
Norris, J.E., \& Da Costa, G.S. 1995, ApJ, 447, 680

\bibitem[2003]{or03}  
Origlia, L., Ferraro, F.R., Bellazzini, M., \& Pancino, E. 2003, 
ApJ, 591, 916

\bibitem[2000]{p00}   
Pancino, E., Ferraro, F.R., Bellazzini, M., et al. 2000, ApJ, 583, L83 

\bibitem[2002]{p02}   
Pancino, E., Pasquini, L., Hill, V., Ferraro, F.R., \& Bellazzini, M. 
2002, ApJ, 568, L101

\bibitem[2003]{p03}
Pancino, E., Origlia, L., Ferraro, F.R., et al. 2003, ASPC, 296, 226

\bibitem[2010]{p10a}   
Pancino, E., Carrera, R., Rossetti, E., \& Gallart, C. 2010a, A\&A, 511, 56 

\bibitem[2010]{p10b}  
Pancino, E., Rejkuba, M., Zoccali, M., \& Carrera, R. 2010b, A\&A, 524, 44

\bibitem[2005]{p05}   
Pasquini, L., Bonifacio, P., Molaro, P., et al. 2005, A\&A, 441, 549

\bibitem[1993]{pet93} 
Peterson, R.C., Dalle Ore, C.M., \& Kurucz, R.L. 1993, ApJ, 404, 333

\bibitem[2008]{pig08} 
Pignatari, M., \& Gallino, R. 2008, AIPC, 990, 336


\bibitem[2004]{pl04}
Plez, B., Hill, V., Cayrel, R., et al. 2004, A\&A, 428, L9

\bibitem[1992]{rai92} 
Raiteri, C. M., Gallino, R., \& Busso, M. 1992, ApJ, 387, 263

\bibitem[1993]{rai93} 
Raiteri, C.M., Busso, M., Neuberger, D., \& K\"{a}ppeler, F. 1993, 
ApJ, 419, 207

\bibitem[2006]{rei06} 
Reijns, R. A., Seitzer, P., Arnold, R., et al. 2006, A\&A, 445, 503 

\bibitem[2009]{ro09}  
Roederer, I.U., Kratz, K.-L., Frebel, A., et al. 2009, ApJ, 698, 1693

\bibitem[2011]{ro11}
Roederer, I.U. 2011, ApJ, 732, L17

\bibitem[1932]{rg32}  
Rose, J.L., \& Granath, L.P. 1932, Phys. Review, 40, 760

\bibitem[1995]{sl95}  
Sarajedini, A., \& Layden, A.C. 1995, AJ, 109, 1086

\bibitem[2011]{s10}   
Shen, Z.-X., Bonifacio, P., Pasquini, L., Zaggia, S. 2010, A\&A, 524, L2

\bibitem[2004]{sim04} 
Simmerer, J., Sneden, C., Cowan, J.J., et al. 2004, ApJ, 617, 1091

\bibitem[2000]{sm00}  
Smith, V.V., Suntzeff, N.B., Cunha, K., et al. 2000, AJ, 119, 1239 (S00)

\bibitem[1983]{sn83}
Sneden, C. \& Parthasarathy, M. 1983, ApJ, 267, 757

\bibitem[1997]{sn97} 
Sneden, C., Kraft, R.~P., Shetrone, M.~D., Smith, G.~H., Langer, G.~E., 
\& Prosser, C.~F.\ 1997, AJ, 114, 1964

\bibitem[2011]{so11} 
Sobeck, J.~S., et al.\ 2011, AJ, 141, 175

\bibitem[2005]{sol05a} 
Sollima, A., Ferraro, F.R., Pancino, E., Bellazzini, M. 2005a, MNRAS, 357, 265

\bibitem[2005]{sol05b} 
Sollima, A., Pancino, E., Ferraro, F.R., et al. 2005b, ApJ, 634, 332

\bibitem[2007]{sol07} 
Sollima, A., Ferraro, F.R., Bellazzini, M., et al. 2007, ApJ, 654, 915 

\bibitem[1981]{stets} 
Stetson, P.B. 1981, AJ, 86, 687

\bibitem[2007]{the07} 
The, L.-S., El Eid, M.F., \& Meyer, B.S. 2007, ApJ, 655, 1058

\bibitem[2004]{tr04}  
Travaglio, C., Gallino, R., Arnone, E., et al. 2004, ApJ, 601, 864 

\bibitem[1981]{truran}
Truran, J.W. 1981, A\&A, 97, 391

\bibitem[2006]{vdv}   
van de Ven, G., van den Bosch, R.C.E., Verolme, E.K., \& de Zeeuw, P.T. 
2006, A\&A, 445, 513 

\bibitem[2000]{vl00}  
van Leeuwen, F., Le Poole, R.S., Freeman, K.C., \& de Zeeuw, P.T. 2000, 
A\&A, 360, 472

\bibitem[2004]{venn}  
Venn, K.A., Irwin, M., Shetrone, M.D., et al. 2004, AJ, 128, 1177 

\bibitem[2009]{vd09}  
Ventura, P., D'Antona, F., Mazzitelli, I., \& Gratton, R.G. 2001, 
ApJ, 550, L65

\bibitem[2010]{vil10} 
Villanova, S., Geisler, D., Piotto, G. 2010, ApJ, 722, L18

\bibitem[1966]{w66}   
Woolley, R.R. 1966, R. Obs. Ann, 2, 1

\bibitem[2008]{yo08}  
Yong, D., Grundahl, F. 2008, ApJ, 672, L29


\end{thebibliography}
\end{document}